\documentclass[aps,prb,amsmath,twocolumn,amssymb,floatfixng,showpacs,
superscriptaddress,footinbib]{revtex4-1}
\pdfoutput=1
\usepackage{float}
\usepackage[dvips]{graphics}
\usepackage{bm}
\usepackage{epsfig}
\usepackage{enumerate}
\usepackage{subfigure}
\usepackage{color}
\usepackage{graphicx}
\usepackage{hyperref}
\hypersetup{
    pdfnewwindow=true,      
    colorlinks=true,       
    linkcolor=magenta,          
    citecolor=blue,        
    filecolor=blue,      
    urlcolor=blue        
}

\newcommand{\ie}{{\it i.e., }}

\newcommand{\ra}{{\rangle }}
\newcommand{\eps}{{\epsilon }}
\newcommand{\bea}{\begin{eqnarray}}
\newcommand{\eea}{\end{eqnarray}}
\newcommand{\beq}{\begin{equation}}  
\newcommand{\eeq}{\end{equation}}

\begin{document} 
\title{Driven conductance of an irradiated semi-Dirac material} 
\author{SK Firoz Islam}
\email{firoz@iopb.res.in}
\affiliation{Institute of Physics, Sachivalaya Marg, Bhubaneswar-751005, India}
\author{Arijit Saha}
\email{arijit@iopb.res.in}
\affiliation{Institute of Physics, Sachivalaya Marg, Bhubaneswar-751005, India}
\affiliation{Homi Bhabha National Institute, Training School Complex, Anushakti Nagar, Mumbai 400085, India}
\begin{abstract}
We theoretically investigate the electronic and transport properties of a semi-Dirac material under the influence of
an external time dependent periodic driving field (irradiation) by means of Floquet theory.
We explore the inelastic scattering mechanism between different side-bands, induced by irradiation,
by using Floquet scattering matrix approach. The scattering probabilities between two nearest side-bands
depend monotonically on the strength of the amplitude of the irradiation. The external irradiation
induces gap into the band dispersion which is strongly dependent on the angular
orientation of momentum. Although, the high frequency limit indicates that the gap opening
does not occur in an irradiated semi-Dirac material, a careful analysis of the full band structure
beyond this limit reveals that gap opening indeed appears for higher values of momentum (away from the Dirac point). 
Furthermore, the angular dependent dynamical gap is also present which cannot be captured within the high frequency approximation.
The contrasting features of irradiated semi-Dirac material, in comparison to irradiated graphene,
can be probed via the behavior of conductance. The latter exhibits the appearance of non-zero conductance dips
due to the gap opening in Floquet band spectrum. Moreover, by considering a nanoribbon geometry of such material,
we also show that it can host a pair of edge modes which are fully decoupled from the bulk, which is in contrast 
to the case of graphene nanoribbon where the edge modes are coupled to the bulk. We also investigate that if the 
nanoribbon of this material is exposed to the external irradiation, decoupled edge modes penetrate into the bulk.
\end{abstract}
\maketitle
\section{Introduction}
In recent times, the research interests on externally irradiated two dimensional (2D) Dirac material
have grown significantly in many aspects among the research community. One of the remarkable 
aspect of this study lies in the  ability of irradiation to modulate the electronic band structure
and most importantly to generate topologically protected conducting edge modes with insulating bulk
in 2D electronic systems (Floquet topological insulator)~\cite{PhysRevB.79.081406,lindner2011floquet,PhysRevB.85.125425,
PhysRevB.84.235108,PhysRevB.85.205428,PhysRevB.89.235416,PhysRevB.94.081103,ezawa2013photoinduced,
PhysRevB.89.121401,cayssol2013floquet}. The  experimental observations of
light induced topological insulating phase have also been reported~\cite{peng2016experimental,zhang2014anomalous,
wang2013observation}. The importance of irradiation effects on 2D Dirac materials are not only limited to 
engineering the Floquet topological insulator, in fact it has also been proposed to modulate spin and valley
degree of freedom in graphene~\cite{PhysRevB.84.195408}, $\alpha$-$T_3$ lattice~\cite{tarun}, 
$\rm MoS_2$~\cite{PhysRevB.90.125438} and emergence of Floquet edge states in germenene~\cite{tahir2016floquet}.
Among the other applications of periodic driving, tuning the thermoelectric performance in
$\rm MoS_2$~\cite{tahir2014tunable} and thin topological insulator~\cite{PhysRevB.91.115311}, 
$0$-$\pi$ phase transition in Josephson junctions of silicene~\cite{PhysRevB.94.165436}
and Weyl semi-metal~\cite{PhysRevB.95.201115} etc. are also worth to be mentioned.

\begin{figure}[!thpb]
\centering
\includegraphics[height=2.5cm,width=0.5 \linewidth]{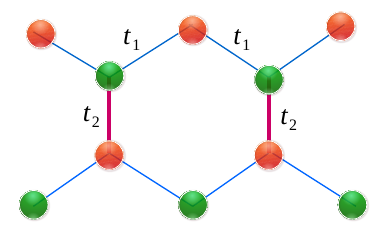}
\caption{(Color online) The hexagonal lattice geometry of a semi-Dirac material with different hoping parameters
$t_1$ and $t_2$ is illustrated. In our case, the hopping amplitude between two nearest zigzag chains is two times
stronger than the other i.e., $t_2=2t_1$. Two different colors are used to denote 
different strength of hopping parameters. On the other hand, two different colored bubbles (light red and light green)
represent two sublattices.}
\label{lattice}
\end{figure}

Along the same line, several attempts have also been made to look into the electronic properties of semi-Dirac material
under the influence of either circularly or linearly polarized light, in the form of irradiation~\cite{PhysRevB.94.081103,PhysRevB.91.205445,PhysRevB.97.035422}.
The Semi-Dirac materials exhibit quadratic as well as linear band dispersion along two orthogonal
momentum directions~\cite{banerjee2009tight,dietl2008new,PhysRevB.97.035422}. The lattice structure of this material
mimics the monolayer graphene with two different hopping parameters $t_1$ and $t_2$, as shown in Fig.~\ref{lattice}.
This kind of lattice has been fabricated into the three unit cell slab of $\rm VO_2$ confined within insulating $\rm TiO_2$.
A special interest has been shown to the case of $t_2=2t_1$ which manifests the semi-Dirac band dispersion
in the vicinity of single Dirac cone. By using an effective low energy Hamiltonian,
a series of theoretical studies have been performed like collective excitations (plasmon)~\cite{PhysRevB.93.085145},
irradiation effects on electronic bulk band structure~\cite{PhysRevB.94.081103,PhysRevB.91.205445},
optical properties~\cite{PhysRevB.97.035422,ales}, disorder effects~\cite{sriluckshmy2018interplay},
interaction effects\cite{interaction}, Landau-Zener oscillations\cite{PhysRevB.96.045424}
and Landau level formation in some specific geometry~\cite{kush_arxiv} etc.

Unlike the graphene case where external irradiation can open up a gap~\cite{PhysRevB.79.081406}, the Floquet band 
structure of the semi-Dirac material remains gapless within high frequency limit as reported in
Ref.~[\onlinecite{PhysRevB.91.205445}]. On the other hand, a topological phase transition can take place
under the consideration of a Haldane type mass term in such material, as predicted in Ref.~[\onlinecite{PhysRevB.94.081103}].
These are the two works which have recently addressed the issue of how irradiation affects the electronic band structure
within the high frequency approximation. 

In this article, we study the quantum transport properties of a semi-Dirac material, exposed
to a circularly polarized light. By using Floquet scattering matrix formalism~\cite{PhysRevB.60.15732}, we compute 
different scattering amplitudes and show that the transmission probabilities through different Floquet side 
bands are monotonically increasing with the strength of the irradiation instead of frequency.
Although it was pointed out earlier that the irradiation cannot open a gap in semi-Dirac material~\cite{PhysRevB.91.205445}
within the high frequency limit, we reveal in our work that beyond this limit a gap indeed opens up in the Floquet quasi-energy spectrum
for non-zero values of momentum \ie~away from the Dirac point. Moreover, unlike the graphene case where the gap parameter 
is a constant mass term~\cite{PhysRevB.79.081406}, it is not constant in semi-Dirac material. Rather, it depends on
the angular orientation of the momentum. Furthermore, several dips appear in the angle averaged conductance spectrum
which are in fact due to the effect of irradiation originating from the higher order correction to
the Floquet spectrum. The earlier prediction that no gap appears under the influence of an irradiation
within high frequency limit, is true only for particular angular orientation of momentum which is 
also confirmed through non-zero conductance dip at undoped situation. This is in contrast to the
case of graphene where zero conductance sharp dip appears. This is due to the fact that external irradiation can induce
a momentum independent constant mass term in graphene, which yields a sharp conductance dip~\cite{PhysRevB.96.245404}.

We also show that the zigzag nanoribbon of this material, without any irradiation, exhibits a pair
of decoupled edge modes which is unlikely to appear in monolayer graphene where edge modes merge
into the bulk at two valleys. However, if the ribbon is exposed to the irradiation these edge modes
no longer remain fully separated from the bulk. Finally, we also carry out the analysis of conductance behavior
through the nanoribbon geometry of an irradiated semi-Dirac material. 
We observe similar conductance dips. However, the overall amplitude of the conductance remains less sensitive 
to the location of the chemical potential, except at the dips. This arises due to the very slow variation of the number of 
transverse edge modes crossing through the chemical potential.

The remainder of the paper is organized as follows. In Sec.~\ref{sec2}, we introduce the low energy effective 
Hamiltonian and corresponding band structure with and without the presence of external irradiation.
The Floquet scattering formalism and conductance calculation for the bulk semi-Dirac material are
presented in Sec.~\ref{sec3}. On the other hand, the Sec.~\ref{sec4} is devoted to the discussion of band spectrum and
conductance features of a nanoribbon geometry both in absence and presence of the periodic driving.
Finally, we summarize and conclude in Sec.~\ref{sec5}.
\section{Model Hamiltonian and band dispersion}\label{sec2}
\subsection{Non-irradiated case}
In this subsection, we discuss the basic features of the band structure when the system is not exposed
to the external irradiation. We begin with the single particle low energy effective model Hamiltonian, based on the 
tight binding model description~\cite{PhysRevB.94.081103,banerjee2009tight,dietl2008new}, as
\begin{equation}
 H_0=\frac{p_y^2}{2m^{*}}\sigma_x+v_{_F}p_x\sigma_y\ ,
\end{equation}
where the effective mass term $m^{\ast}$ and the velocity $v_F$ can be expressed explicitly in terms of
tight-binding model parameters as $v_F=3ta/\hbar$ and $m^{*}=2\hbar/(3ta^2)$. Here, $a$ and $t=t_{1}$ are the lattice constant 
and one of the hopping parameters, respectively. The Pauli matrices ${\bf\sigma}\equiv\{\sigma_x,\sigma_y\}$ denote
pseudo sublattice spin index. The $2$D momentum operator is denoted by ${\bf p}\equiv\{p_x,p_y\}$.
\begin{figure}[!thpb]
\centering
\includegraphics[height=5cm,width=0.8 \linewidth]{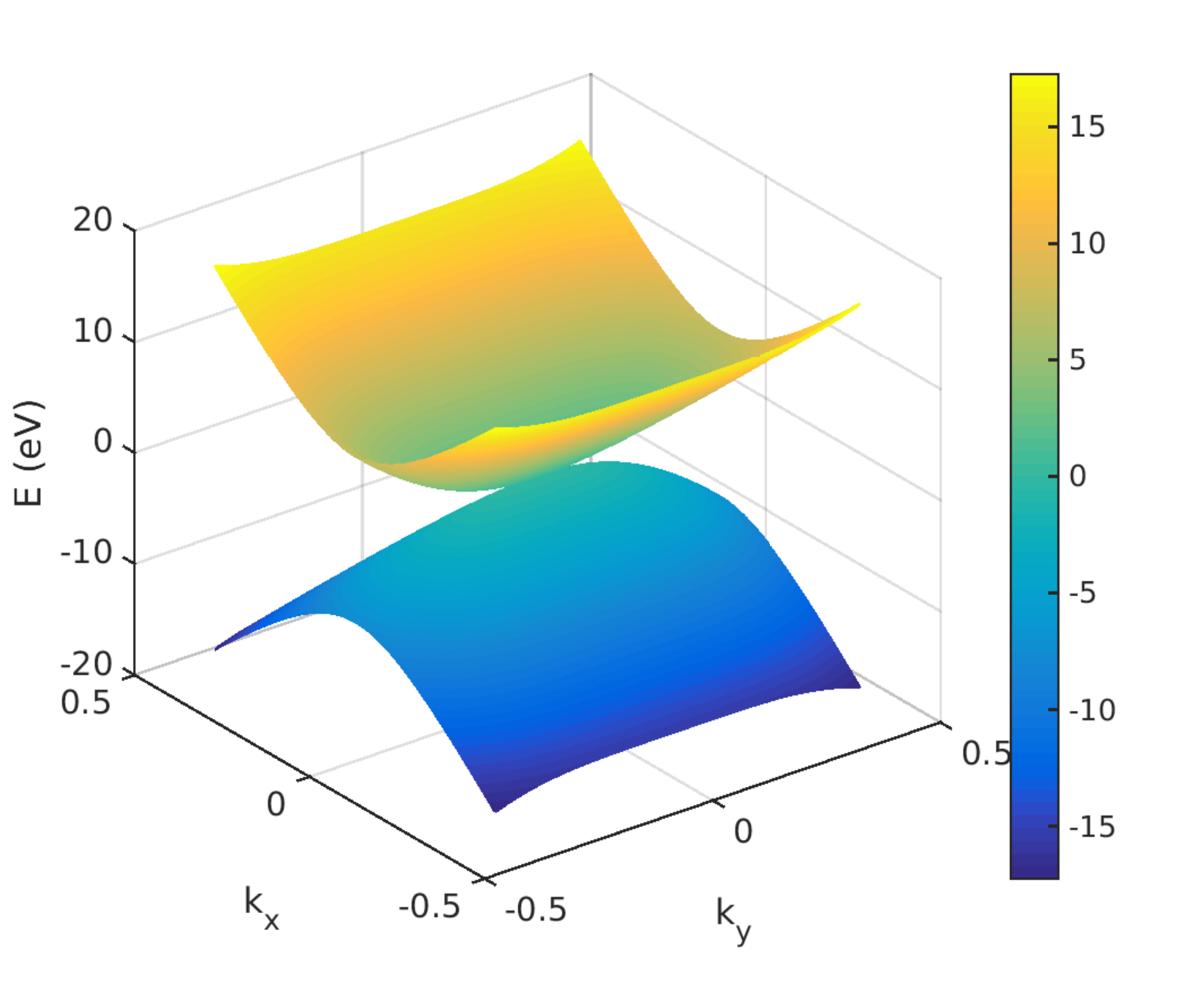}
\caption{(Color online) Anisotropic energy band dispersion of a semi-Dirac material is shown for the non-irradiated case. 
Here the momentum axes are normalized by $k_0=10^{10}m^{-1}$.}
\label{band_dis}
\end{figure}
The energy band dispersion of this semi-Dirac material can be obtained as
 \begin{equation}\label{band}
  E_{k_x,k_y}=\lambda\sqrt{(\hbar v_{_F} k_x)^2+\left(\frac{\hbar^2 k_y^2}{2m^{\ast}}\right)^2},
 \end{equation}
where $\lambda=\pm$ is the band index and the $2$D momentum vector is given by ${\bf k}=\{k_x,k_y\}$.
Note that, the energy dispersion is quadratic along $k_y$ direction and linear along $k_x$ direction,
justifying the nomenclature of such materials as semi-Dirac material. A contour plot of the band dispersion
is demonstrated in Fig.~\ref{band_dis} which indicates the anisotropic band structure of such material.
\begin{figure*}
 {\includegraphics[width=.33\textwidth,height=5cm]{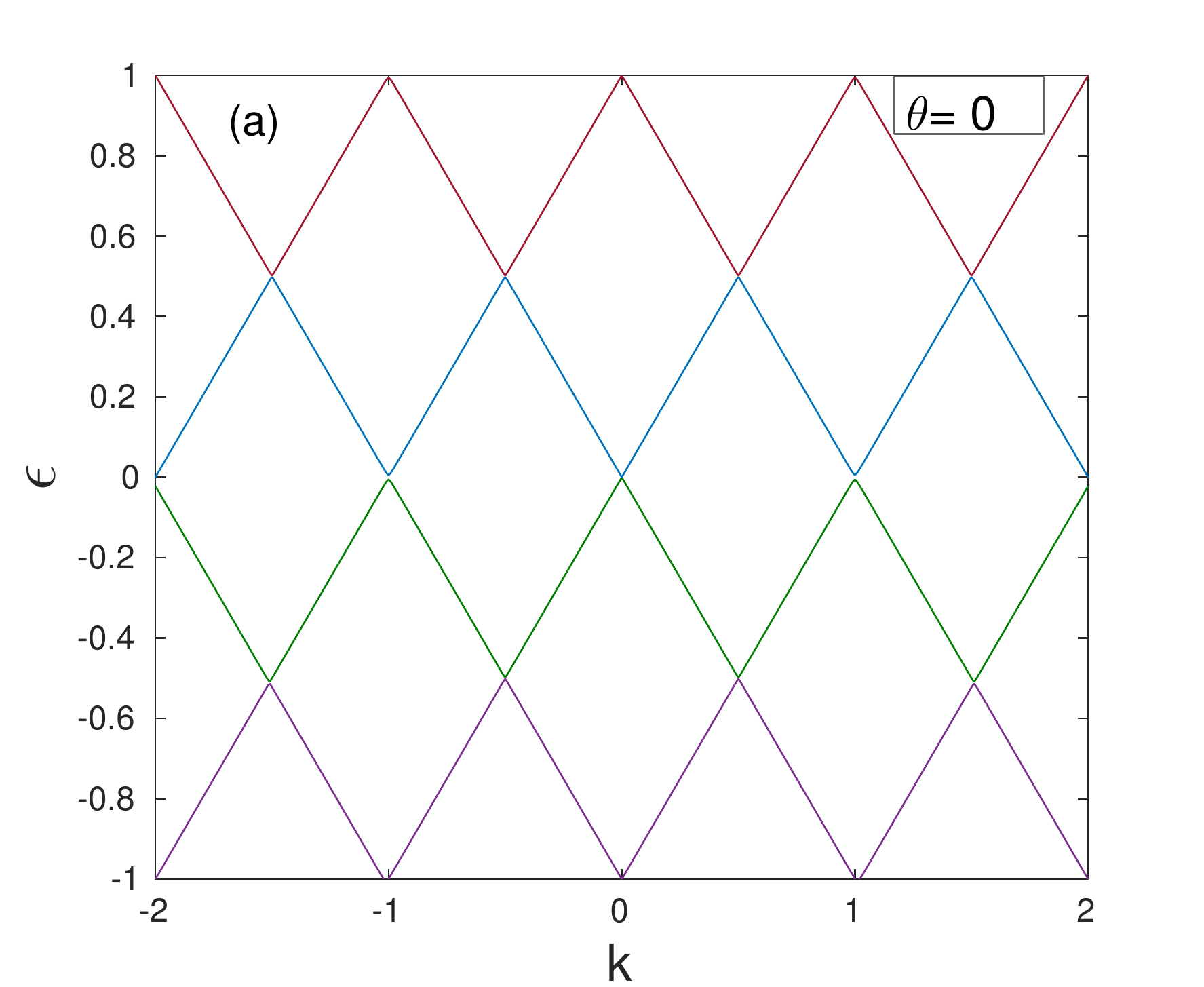}}
 \hspace{-2em}
 {\includegraphics[width=.33\textwidth,height=5cm]{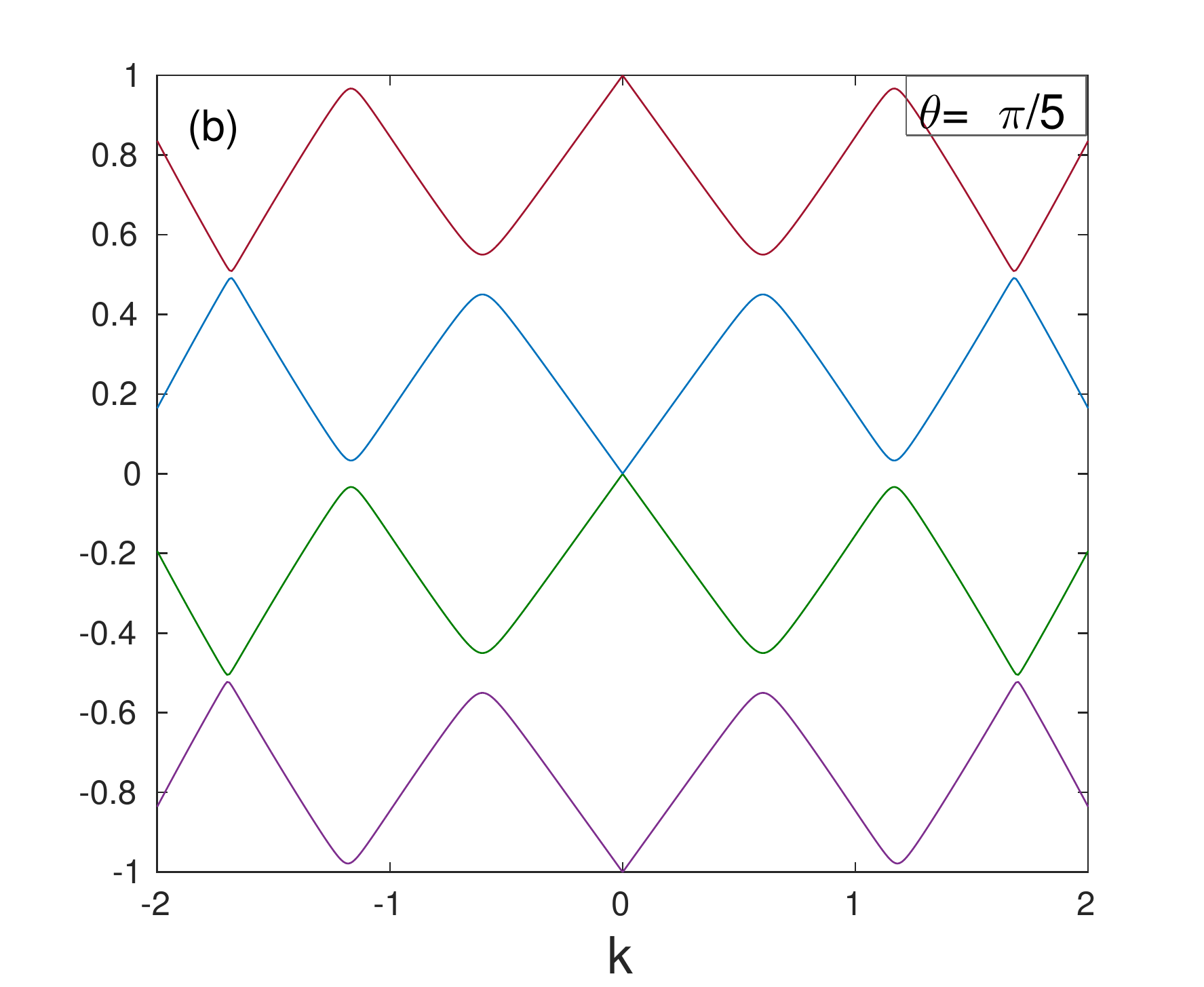}}
 \hspace{-2em}
 {\includegraphics[width=.33\textwidth,height=5cm]{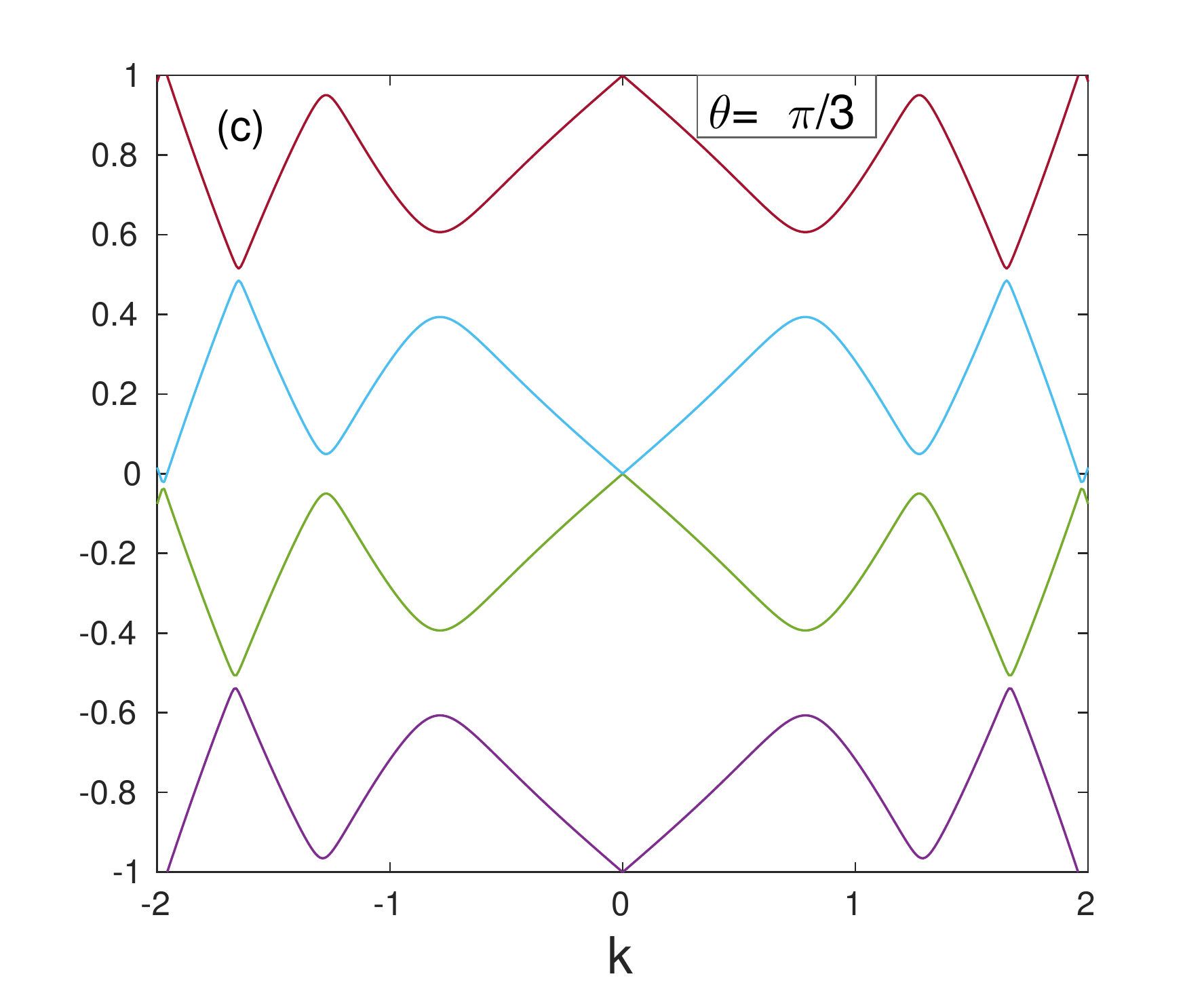}}
\caption{Floquet quasi energy spectrum is illustrated as a function of $k~(=\sqrt{k_{x}^{2} + k_{y}^{2}})$ for three different angular orientation 
of momentum $\theta~(=\tan^{-1}(k_y/k_x)$): (a) $\theta=0$ (b) $\theta=\pi/5$ and (c) $\theta=\pi/3$. The quasi energy and wave vector are 
in units of $\hbar\omega$ and $l_\omega^{-1}=\omega/v_F$, respectively. Here, we consider the irradiation strength to be $\beta=0.3$.}
\label{quasi_energy}
\end{figure*}
\subsection{Irradiated case}
This subsection is devoted to address the effect of irradiation on the band structure of the semi-Dirac material via the
low energy effective Hamiltonian. We consider a time dependent periodic perturbation as irradiation under which 
the dynamics of the charge carriers can be described by the Floquet theory~\cite{RevModPhys.89.011004}. 
This is analogous to the Bloch's theorem about the electronic motion in presence of a space dependent periodic potential.
The irradiation is considered to be a circularly polarized light which can be expressed as a vector 
potential ${\bf A}=A_0[\cos(\omega t)\hat{x}+\sin(\omega t) \hat{y}]$. Here, $\omega$ and $A_0$ are the frequency 
and amplitude of the irradiation, respectively. To include this vector potential into the low energy effective Hamiltonian,
we use the Peiere's substitution ${\bf p\rightarrow (p+eA)}$ and obtain the irradiated Hamiltonian as
\begin{equation}
 H_{T}=H_{0}+\mathcal{H}_{1}\sigma_{x}+\mathcal{H}_{2}\sigma_y\ ,
\end{equation}
where 
\begin{equation}
 \mathcal{H}_1=\left[Fk_y\cos(\omega t)+G\{1-\cos(2\omega t)\}\right]\ .
\end{equation}
and
\begin{equation}
 \mathcal{H}_{2}=R\sin(\omega t)\ .
 \end{equation}
 with $F=eA_0/m^{*}$, $G=(eA_0/2m^{*})^2$ and $R=eA_0v_{_F}$.

 To solve the above Hamiltonian, we employ the Floquet formalism which  infers that the system
 under the time dependent periodic perturbation exhibits a complete set of 
 solutions of the form $\Psi({\bf r},t)=\phi({\bf r,t})\exp(-i\eps t/\hbar)$, where 
 $\lvert\phi({\bf r},t+T)\ra=\lvert\phi(\bf r,t)\ra$ is the corresponding Floquet
states and $\eps$ is the quasi energy~\cite{RevModPhys.89.011004}. Here, $T$ is the periodicity of the 
perturbation. Now, we can directly substitute the Floquet eigen states into the 
time dependent $\rm Schr\ddot{o}dinger$ equation and obtain a time independent $\rm Schr\ddot{o}dinger$ type equation
$\hat{H}_{F}\phi({\bf r, t})=\eps\phi({\bf r, t})$ with the 
Floquet Hamiltonian $\hat{H}_{F}=\hat{H}_0-i\hbar(\partial/\partial t)$. The Floquet eigen states
can be further expressed in the Fourier expansion as~\cite{RevModPhys.89.011004}
\begin{equation}
 \phi({\bf r, t})=\sum_{n}\phi_{n}({\bf r,t})e^{in\omega t}\ .
\end{equation}
where $n$ denotes the Floquet modes or side bands. Hereafter, we discuss the Floquet
band structure of the irradiated semi-Dirac materials in two different regimes of frequency, $\omega$.
\subsubsection{High frequency limit:}
Here, we briefly review the band structure of the semi-Dirac material within the high frequency limit
which has been previously considered in Ref.~[\onlinecite{PhysRevB.91.205445}].
In this limit, the irradiated Hamiltonian can be recast to an effective Hamiltonian as~\cite{PhysRevB.91.205445} 
\begin{equation}\label{eff}
 H_{eff}\simeq H_0+\frac{[H_{-1},H_{+1}]}{\hbar\omega}\ .
\end{equation}
where the second term arises due to the irradiation and
\begin{equation}
H_{m}=\frac{1}{T}\int_{0}^{T}dt e^{-im\omega t}H(t)\ .
\end{equation}
with $H_{+1}=H_{-1}^{\dagger}=(Fk_y\sigma_x-iR\sigma_y)/2$. Incorporating these terms in Eq.(\ref{eff}),
the effective Hamiltonian can be found to acquire a direction dependent mass term as $-\eta[(eA_0)^2v_Fk_y/(m^{*}\hbar\omega)]
\sigma_z$ with $\eta=+(-)$ denotes left (right) circularly polarized light. The corresponding energy eigen value can be subsequently
written as
\begin{equation}
 \eps=\pm\sqrt{\left(\frac{k_y^2}{2m^{*}}\right)^2+(\hbar v_F k_x)^2+\left(\eta\frac{e^2A_0^2v_Fk_y}{m^*\hbar\omega}\right)^2}.
\end{equation}
Note that, the mass term vanishes at $k_{y}=0$.
\ie irradiation cannot open up a gap in the Floquet band structure. This was in fact the main claim
of Ref.~[\onlinecite{PhysRevB.91.205445}]. However, we emphasize here that a higher order term ($G=[eA_0/2m^{*}]^2$) 
in the total irradiated Hamiltonian has been ignored.

Note that the high frequency results, discussed in Ref.~[\onlinecite{PhysRevB.94.081103}], is presented for a 
photoinduced vector field of the form $A=A_0[\sin(\omega t), \sin(\omega t+\phi)]$.
To recover the result of circularly polarized light, one need to set $\phi=\pi/2$.

\subsubsection{Beyond the high frequency limit:}
In this subsection, we investigate the full band structure of the semi-Dirac material beyond the high frequency approximation.
Using the Fourier expansion of Floquet modes, we can write the time-independent eigen value problem as 
\begin{equation}\label{eveq}
 \sum_{n'}H_{F,nn'}\phi_{n}=\sum_{n'}[H_{0F,nn'}+\mathcal{V}_{F,nn'}]\phi_{n}=\eps \phi_{n}\ .
\end{equation}
where the diagonal part of the Floquet Hamiltonian can be defined as
\begin{equation}
H_{0F,nn'}=\left[\frac{p_y^2}{2m^{\star}}\sigma_x+v_{_F}p_x\sigma_y-\mu+n'\hbar\omega\right]\delta_{nn'}\ .
\end{equation}
and the irradiation induced coupling Hamiltonian between different Floquet side-bands can be written as
\begin{eqnarray}
 \mathcal{V}_{F,nn'}&&=[H_{+1}\delta_{n,n'-1}+H_{-1}\delta_{n,n'+1}]\nonumber\\&&
 +[H_{+2}\delta_{n,n'-2}+H_{-2}\delta_{n,n'+2}] \ .
\end{eqnarray}
Here, $H_{+2}=H_{-2}^{\dagger}=(eA_0)^2/4m^{*}$.
Note that, to diagonalize the Floquet eigen value problem, we need to set a cut-off to the dimension of the Floquet 
space. For example, if we impose the cut-off to the Floquet side band at $N$, then the size 
of the matrix becomes $2(2N+1)\times 2(2N+1)$ dimensional including the sublattice degree of freedom $2$.
In our present case, we consider $n=2$ for the analysis of quasi-energy and transport properties.
This is a valid approximation, as the higher side-bands cause vanishingly small contribution to the transport.
We show the quasi energy band spectrum in Fig.~\ref{quasi_energy} as a function of momentum $k=\sqrt{k_{x}^{2} + k_{y}^{2}}$ 
($k_{x}=k\cos\theta, k_{y}=k\sin\theta$) for three different angular orientation of momentum $\theta=\arctan(k_y/k_x)$, 
which is obtained by numerically diagonalizing the Floquet Hamiltonian $H_{F,nn'}$ (see Eq.(\ref{eveq})).
Also note that, the eigen value equation [Eq.(\ref{eveq})] is normalized by $\hbar\omega$ in order to diagonalize
the Hamiltonian. This introduces two new parameters: $\beta=ev_FA_0/\hbar\omega$ defining the strength of the irradiation
and a length scale $l_{\omega}=v_{F}/\omega$. A finite value of $\beta$ is required for gap opening in the spectrum.
The quasi energy spectrum is gapless for $\theta=0$, whereas a finite gap appears for $\theta\ne 0$.
However, note that even for $\theta\ne 0$, gap opening does not occur at $\epsilon=0$ for $k=0$ (Dirac point).
This was also predicted earlier via high frequency approximation~\cite{PhysRevB.91.205445}.
Another noticeable point here is that unlike the case of irradiated monolayer
graphene~\cite{PhysRevB.96.245404} where gap opening occurs at the momentum values 
$k=\pm j\omega/2v_{F}$ ($j=1,2,3..$), here in semi-Dirac materials it depends on the angular orientation of the momentum.

We would like to emphasize that the numerical results for the Floquet band structure (in the intermediate frequency regime), presented in Fig.~\ref{quasi_energy}, 
manifest that no gap opens at $k=0$ which was also predicted via the high frequency approximation in Ref.~[\onlinecite{PhysRevB.91.205445}]. 
However, note that the gap opening away from the Dirac point and other dynamical gaps which appear at $\epsilon=\pm\hbar\omega/2$, cannot be 
captured via the high frequency approximation.

\section{Methodology and numerical results for the bulk transport}\label{sec3}
\begin{figure}[!thpb]
\centering
\includegraphics[height=6cm,width=0.95 \linewidth]{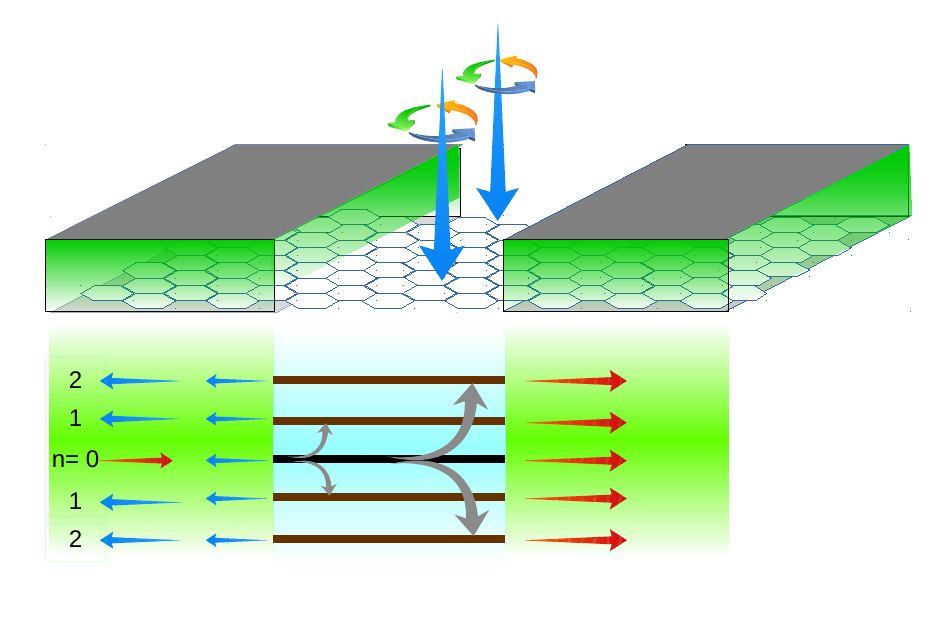}
\caption{(Color online) A schematic sketch of the device, composed of a semi-Dirac material, is demonstrated in
the upper panel and scattering mechanism is shown in the lower panel of the figure. The central region of the above device 
is illuminated by circularly polarized light. The central region is attached to two reservoirs ($x=0$ and $x=L$) which are also comprised 
of the semi-Dirac material. The left and right regions are protected from any kind of effects of irradiation by placing 
a wall (grey color) on it.}
\label{device}
\end{figure}

In this section, we discuss the quantum transport properties through the irradiated bulk region of the semi-Dirac
material by using Floquet scattering matrix formalism~\cite{PhysRevB.60.15732}. The relevant geometry is schematically 
depicted in Fig.~\ref{device}. The set up is composed of three different regions in which left and right regions are protected from the 
exposure of irradiation by placing a cover on it (as shown by grey color) while the middle region is illuminated by 
circularly polarized light.
\subsection{Solving scattering problem using Floquet method}
In order to solve the scattering problem, we employ the approach prescribed in Ref.~[\onlinecite{PhysRevB.60.15732}]
which is very recently employed in graphene~\cite{PhysRevB.96.245404}. The non-irradiated left region is doped in
such way that the chemical potential matches with the zeroth ($n=0$) Floquet side band inside the central
irradiated region~\cite{PhysRevB.96.245404}. In such case, the incoming electron from the left  region will be
scattered inelastically among different Floquet side-bands inside the irradiated region by absorption/emission
of photon process (see the lower panel of Fig.~\ref{device}). Subsequently, the reflection and transmission
processes will occur from all Floquet bands. To proceed further, we need to obtain the Floquet states and
momenta ($k_x$) of electrons scattered from each side band, for a particular $k_y$. To obtain that,
we multiply $\sigma_y$ from the left side in Eq.(\ref{eveq}) and rearranging it we get
\begin{equation}\label{kxx}
\sum_{n}(Q_{0F,nn'}+Q_{VF,nn'})\phi_{n}=\hbar v_{_F}k_x\phi_{n}\ .
\end{equation}
where
\begin{equation}
 Q_{0F,nn'}=\sigma_y\left[(\mu-n'\hbar\omega)-\frac{p_y^2}{2m^{\ast}}\sigma_x\right]\delta_{nn'}\ ,
\end{equation}
and
\begin{equation}
 Q_{V,nn'}=-\frac{\sigma_y}{2}(Fk_y\sigma_x-iR\sigma_y)(\delta_{n,n'-1}+\delta_{n,n'+1}) \ .
\end{equation}
The numerical diagonalization of Eq.~(\ref{kxx}) yields $2(2N+1)$ values of $k_x^{\nu}$
and eigen vectors $\phi_{n}^{\nu}$ with $\nu$ being the index of different eigen modes.
Here, we set the quasi energy at $\epsilon=0$ which corresponds to the incident electron at the
chemical potential of the left normal lead. The wave function for a given $\nu$ can be expressed as
\begin{equation}
 \phi^{\nu}(x,t)\sim e^{ik_{x}^{\nu}x}\sum_{n=-N}^{N}\phi_{n}^{\nu}e^{-in\omega t}\ .
\end{equation}
which is a two component vectors composed of two sub-lattices.
\begin{figure}[h]
\centering
\includegraphics[height=6cm,width=\linewidth]{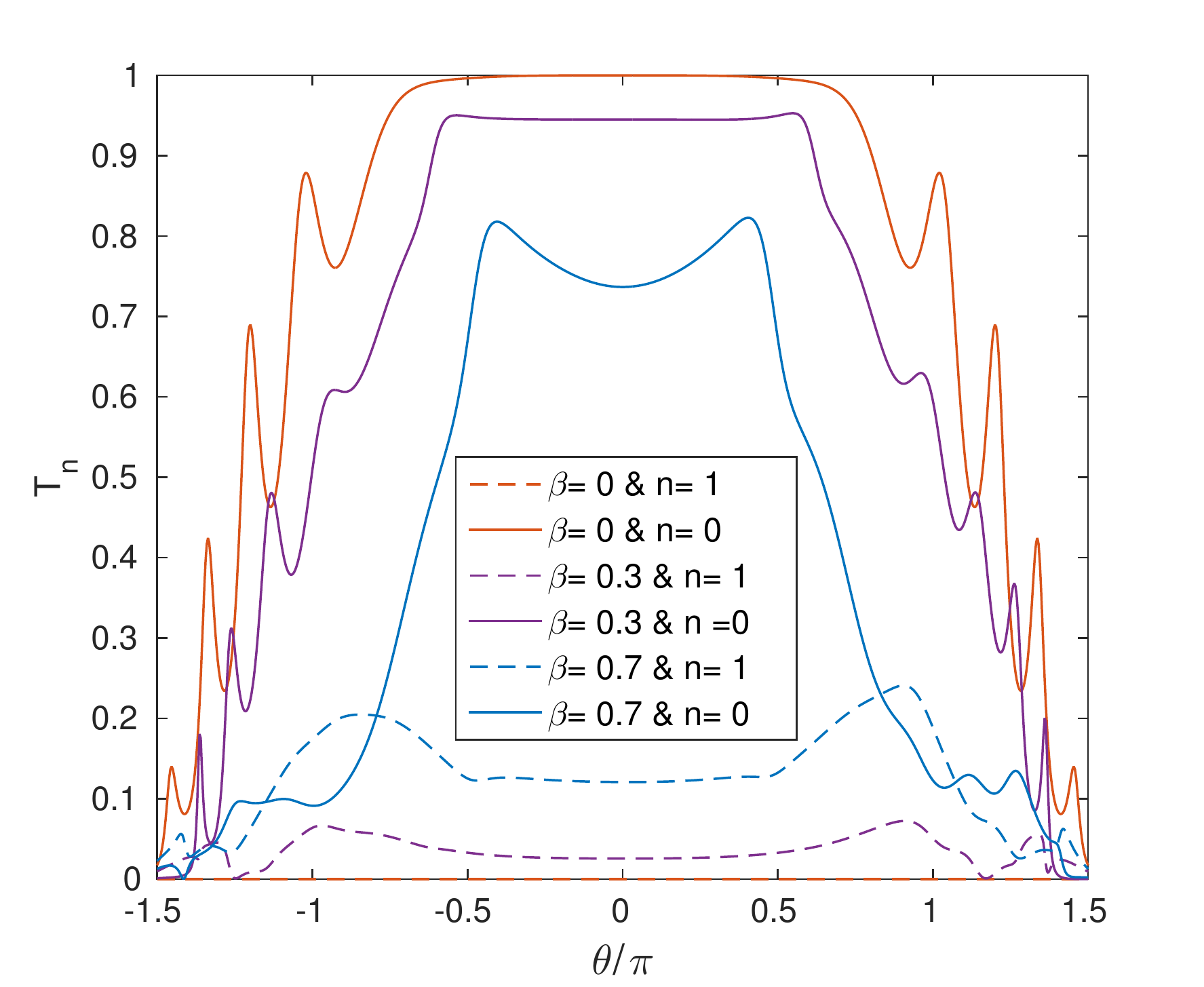}
\caption{(Color online) The behavior of transmission probabilities through different Floquet side bands
are illustrated as a function of angle of incidence in the left normal region (non-irradiated).}
\label{trans}
\end{figure}
Note that, the momentum ($k_x^{\nu}$) can be real or imaginary depending on the location of the chemical potential in the left region.
The imaginary momentum corresponds to the situation when the chemical potential lies inside the gap.
This leads to the emergence of evanescent mode. On the other hand, the spectrum in the left/right non-irradiated region can be obtained 
in similar fashion but with zero strength of the irradiation as those regions are assumed to be free from the external light. However, the electrons 
which are scattered from the irradiated central region to the normal regions via different fictitious Floquet side-bands of the normal regions,  
are governed by the following eigen value equation
\begin{equation}
 \left[\frac{p_y^2}{2m^{\ast}}\sigma_x+v_{_F}p_x\sigma_y\right]\phi_{n}=[\epsilon(\theta)-n\hbar\omega]\phi_{n} \ .
\end{equation}
As the side bands are decoupled from each other in the normal regions due to the absence of irradiation,
the momentum of the scattered electrons in each artificial side band remains identical. Furthermore, the wave function 
of the incident as well as the reflected/transmitted electrons are given by
\begin{equation}
 \phi_{n,\lambda}(x)=\frac{1}{\sqrt{2}}
 \left[\begin{array}{c} \lambda\\ie^{-i\tilde{\theta}}\end{array}\right]e^{ik_{x}x}\  ,
\end{equation}
where $\tan\tilde{\theta}=[\hbar k/(2m^{\ast}v_{_F})]\tan\theta\sec\theta$ with 
$\theta=\tan^{-1}(k_y/k_x)$. After substituting the tight-binding model parameters~\cite{PhysRevB.94.081103} 
for mass and Fermi velocity as $m^{\ast}=2\hbar/(av_{_F})$ and $v_{_F}=3ta/\hbar$,
we obtain $\tan\tilde{\theta}=\tilde{k}\tan\theta/(4\cos\theta)$ with $\tilde{k}=ka$.

\subsection{Boundary conditions at the interfaces}
To solve the scattering problem we match the wave functions at the two interfaces $x=0$ and $x=L$. Here, $L$ is the length of the 
irradiated region. We assume that the electron is incident only from the $n=0$ Floquet band of the left non-irradiated region and 
after encountering inelastic scattering processes inside the irradiated region, it is either reflected or transmitted through all possible 
virtual Floquet side bands. Thus, there will be $(2N+1)$ reflected channels in the left region in which the wave function is given by
\begin{equation}
 \Psi_{L}(x,t)=\left[\phi_{i}e^{ik_{x}x}+\sum_{n=-N}^{N}r_{n0}\phi_{r}e^{-i(k_{x}x+n\omega t)}\right]e^{-i\eps t}\ ,
\end{equation}
\begin{figure*}
 {\includegraphics[width=.49\textwidth,height=6cm]{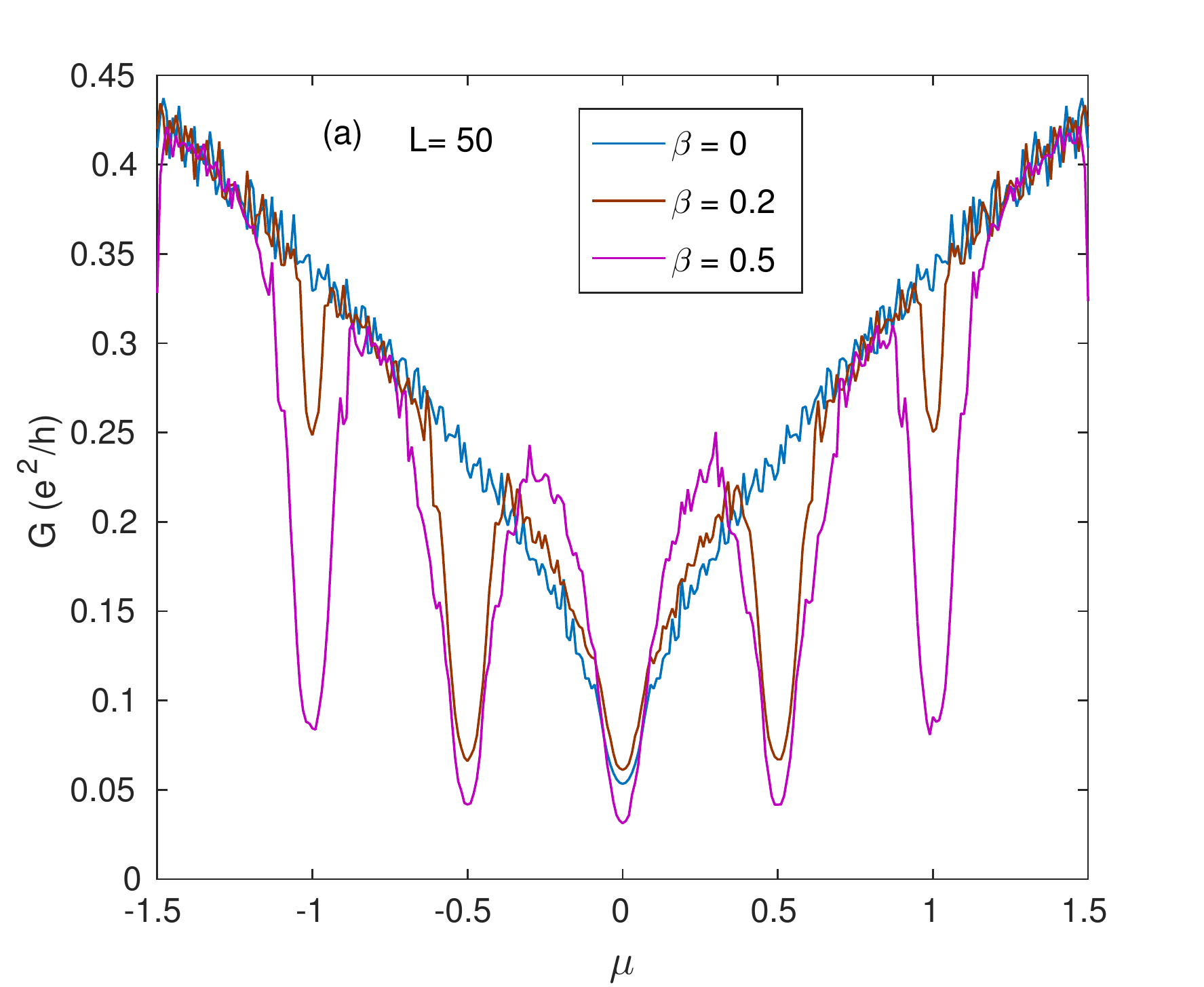}}
 {\includegraphics[width=.49\textwidth,height=6cm]{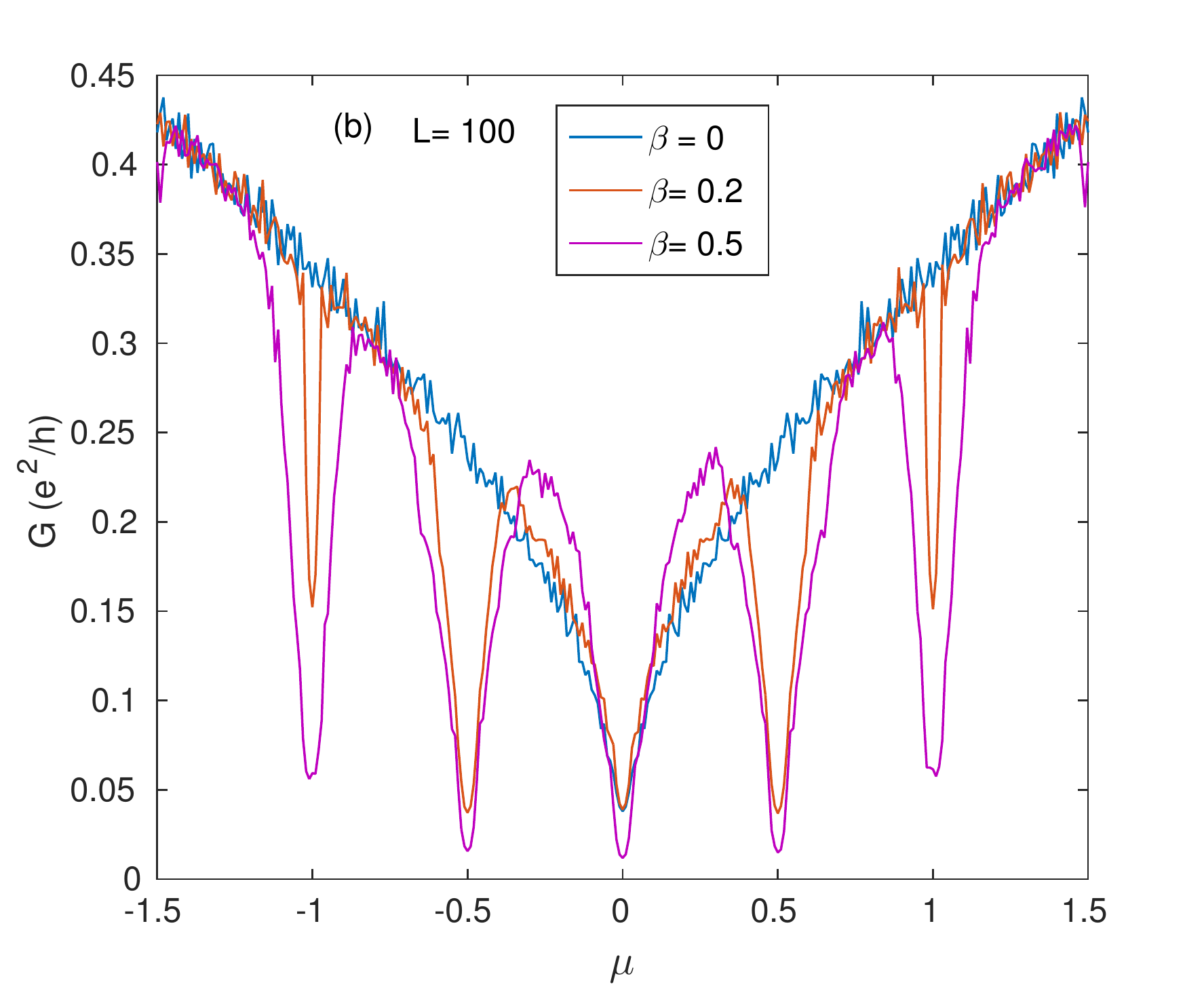}}
\caption{(Color online) Conductance of the periodically driven bulk semi-Dirac material is demonstrated as a function of chemical potential for 
(a) $L=50$ (b) $L=100$. Here, the length and the chemical potential of the irradiated region is normalized by $l_w=v/\omega$ and $\hbar\omega$, respectively.}
\label{cond_plot}
\end{figure*}
where the Floquet coefficients of the incident and reflected waves are
$\phi_{i}=\left[\begin{array}{c}\lambda\\ie^{-i\tilde{\theta}}\end{array}\right]$ and 
$\phi_{r}=\left[\begin{array}{c}\lambda\\ie^{-i\tilde{\theta}_r}\end{array}\right]$ respectively.
Here, $\tan\tilde{\theta}_r=\tan\tilde{\theta}~(\theta\Rightarrow\pi-\theta)$.
The reflection amplitude corresponding to the $n$-th Floquet band is denoted by $r_{n0}$. 
The Floquet eigen states inside the irradiated region can be written as
\begin{eqnarray}
 \Psi_{I}(x,t)=&&\sum_{\nu}a_{\nu}\phi^{\nu}(x,t)e^{-i\eps t}\nonumber\\
 &&=\left[\sum_{\nu}a_{\nu}e^{ik_{x}^{\nu}x}\sum_{n=-N}^{N}\phi_{n}^{\nu}e^{-in\omega t}\right]e^{-i\eps t}\  .
\end{eqnarray}
On the other hand, the wave function in the right non-irradiated region is composed of only transmitted
waves and can be written as
\begin{equation}
 \Psi_{R}(x,t)=\left\{\sum_{n=-N}^{N}t_{n0}\phi_{r}e^{i[k_{x}(x-L)+n\omega t]}\right\}e^{-i\eps t}\ .
\end{equation}
Here, $t_{n0}$ represents the transmission amplitude via $n$-th band into the right normal region.
The reflection and transmission amplitudes can be evaluated by matching the wave functions at the 
two interfaces which yields a set of linear equations as
\begin{equation}
 \phi_{i}\delta_{n0}+r_{n0}\phi_{r}=\sum_{\nu}a_{\nu}\phi_{n}^{\nu}\ ,
\end{equation}
and
\begin{equation}
 \sum_{\nu}a_{\nu}\phi_{n}^{\nu}e^{ik_{x}^{\nu}L}=t_{n0}\phi_{t}\ .
\end{equation}
where $\phi_{t}=\phi_{i}$.
\subsection{Scattering amplitudes and Conductance}
In this subsection, we present the numerical results for ballistic transport through the irradiated semi-Dirac material 
for different strengths of driving. As mentioned earlier, we assume that the chemical potential in the left lead is set 
at such a position that the electrons can only be incident from the virtual $n=0$ Floquet band of the left lead and subsequently 
after entering into the irradiated region it encounters scattering into different Floquet bands via the inelastic scattering 
processes (absorption and emission). The reflection and transmission probabilities for a particular angle of incidence 
and energy of the incident electron are the sum over all reflection and transmission amplitudes via different Floquet bands.
This can be written as
\begin{equation}
 T(\theta,\mu)=\sum_{n=-N}^{n=N}|t_{n0}(\theta,\mu)|^2\ ,
\end{equation}
and
\begin{equation}
 R(\theta,\mu)=\sum_{n=-N}^{n=N}|r_{n0}(\theta,\mu)|^2\ ,
\end{equation}
which satisfies the unitarity condition as $R(\theta)+T(\theta)=1$. Once the transmission amplitudes
are evaluated, one can readily employ the Landauer-Buttiker formula to obtain the conductance by 
averaging over all possible angle of incidence as
\begin{equation}\label{cond}
 G=\frac{e^2}{h}\int_{-\pi/2}^{\pi/2}\sum_{n=-N}^{N}d\theta |t_{n}(\theta,\mu)|^2\ .
\end{equation}

We present the behavior of transmission probabilities via different Floquet side bands in Fig.~\ref{trans}.
Here, the incident electron is considered to be from the zeroth Floquet band ($n=0$) of the non-irradiated left lead.
We consider three Floquet side bands inside the central irradiated region which is sufficient to preserve the probability conservation. 
The transmission probability via $n=1$ side band is shown for different strengths of irradiation $\beta$ (see Fig.~\ref{trans}). 
Note that, $\beta$ couples the two nearest side bands for which the transmission probability increases via $n=1$ side band 
as we enhance $\beta$. This happens due to one photon absorption process. As a result, this reduces the same in $n=0$ side band. 
The incident electron from the left lead at $n=0$ band can be transmitted/reflected through the same band
$n=0$ which we call intra-band scattering whereas transmission/reflection through $n=1$ side band is termed
as inter-side band scattering. The intra-band transmission probability can be $100 \%$ at normal angle of
incidence ($\theta=0$) for $\beta=0$ due to the Klein tunneling. However, with the increase of $\beta$ 
the Klein process becomes suppressed resulting in enhancement of inter-band transmission probability. 
The reason can be attributed to the fact that, as the amplitude of the irradiation increases, the coupling strength 
between the two nearest Floquet side bands increases resulting in enhancement of the inter-band transmission probability. 

The driven conductance of the bulk semi-Dirac material is demonstrated in Fig.~\ref{cond_plot}. Here we use Eq.(\ref{cond}) 
to compute the conductance for two different lengths of the irradiated region.
In Fig.~\ref{cond_plot}(a), the conductance is shown for different strengths of the periodic driving
around chemical potential $\mu=0$ and $\hbar\omega/2$. These correspond to the gaps opening of
the band dispersion, as shown in Fig.~\ref{quasi_energy}. Note that, although the Floquet spectrum is
gapless at $k=0$ which is even confirmed in Ref.~[\onlinecite{PhysRevB.91.205445}] within the 
high frequency approximation, the band gap indeed appears for other values of $k$ (away from the Dirac point) which
yields a dip in the angle averaged conductance at $\mu=0$. We would like to emphasize here that in case of graphene~\cite{PhysRevB.96.245404},
the location of gap appears to be at $k=\pm v_{F}/\omega$. On the contrary, in semi-Dirac material it depends 
on the angular orientation of the momentum as shown in Fig.~\ref{quasi_energy}. Some other conductance
dips also arise at $\mu=\hbar\omega/2$ which correspond to the dynamical gap between two Floquet side bands.
The qualitative features of the driven angle averaged conductance remain similar as we choose a different length of the
central irradiated region (see Fig.~\ref{cond_plot}(b)). However, the conductance dips  become more deeper
due to enhanced probability of Floquet sub-band scattering as the length of the central region increases. 

Note that, the irradiation (circularly polarized light) can open-up a gap in graphene~\cite{PhysRevB.96.245404} 
at $k=0$ (Dirac point) which does not depend on angular orientation of momentum. This can result in rectangular 
shaped sharp conductance dips with almost zero values under the suitable strength of irradiation. On the other hand, 
the driven semi-Dirac material  may not acquire a gap at $k=0$, in the $n=0$ band. However, the conductance dips still appear 
due to the gap opening at other $k$ values away from zero. Although, the conductance cannot be exactly zero in this case 
at $\mu=0$ as the semi-Dirac material still is metallic at $k=0$. Neverthless, around the dynamical gap 
$\mu=\pm\hbar\omega/2,\pm\hbar\omega$, the dips appear due to the gap opening between two nearest side bands.
\section{Band dispersion and Transport through a nanoribbon geometry of semi-Dirac material}\label{sec4}
In this section, we address the electronic and transport properties of this material for the nanoribbon geometry
in presence of periodic driving. However, our discussion for the nanoribbon is restricted to zigzag case only
due to its unique features over armchair one, as demonstrated in case of graphene~\cite{PhysRevB.89.121401}. 
\subsection{Non-irradiated ribbon:}
\begin{figure}[!thpb]
\centering
\includegraphics[height=6cm,width=\linewidth]{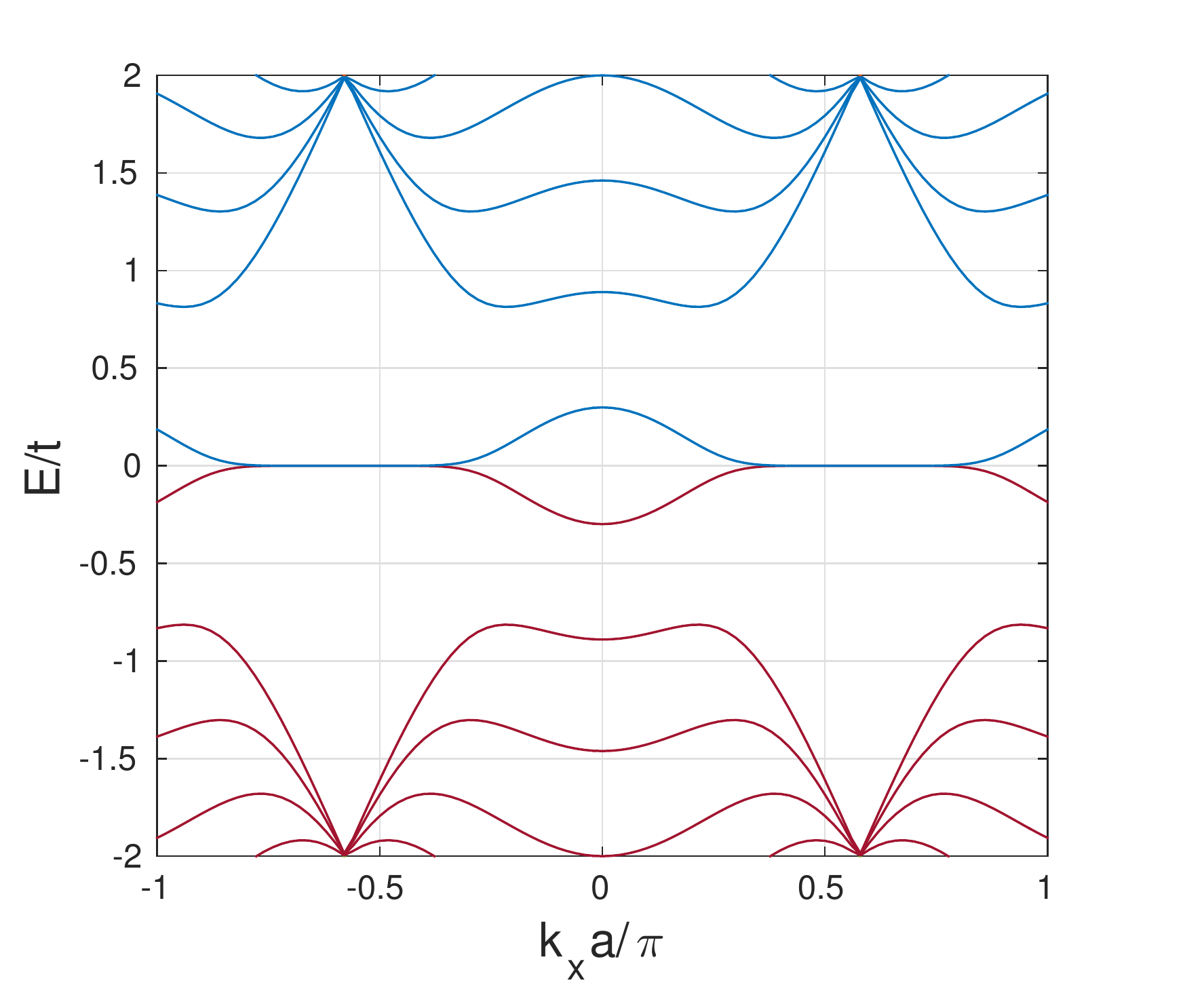}
\caption{(Color online) Energy band dispersion of a semi-Dirac material based ZNR is demonstrated. 
The width of the ribbon is considered as $N=20$.}
\label{ribbo_con}
\end{figure}
In order to find the energy band dispersion of a zigzag nanoribbon of semi-Dirac material,
an effective difference equation can be constructed by employing an analogy to the 
case of an infinite one-dimensional chain~\cite{dutta,PhysRevB.92.035413}.
To implement this, the nanoribbon can be considered as the composition of an array of the 
vertical rectangular unit cells (supercells). The width of the zigzag nanoribbon
(ZNR) is determined by the number of atoms $N$ per unit cell.
The effective difference equation of the ZNR takes the form as
\bea
(E \mathcal{I}-\mathcal{E}) \psi_l=\mathcal{T} \psi_{l+1}+\mathcal{T}^{\dagger} \psi_{l-1}\ ,
\eea
where
\bea
\mathcal{E}=\sum_{\alpha}\sum_{\langle i, j\rangle}t_{\alpha}c^{\dagger}_{i,\alpha}c_{j,\alpha}+h. c\ ,
\label{HNR}
\eea
is the on-site energy matrix with $\alpha=1,2,3$ and
\begin{align}
\psi_l &=\begin{bmatrix} 
\psi_{l,1}\\ \psi_{l,2} \\ \vdots \\ \psi_{l,N} \end{bmatrix}\ ,
\end{align}
Here, $\mathcal{T}$ is the nearest-neighbor hopping matrices of the unit cells.
The vertical supercell~\cite{PhysRevB.92.035413} is denoted by the $l$ index and $\mathcal{I}$ represents 
the unit matrix being of dimension $N \times N$. The zigzag chain of the ribbon is translationally invariant along
the $x$-direction, which yields the momentum ($k_x$) along this direction to be conserved and acts as a good quantum
number. Finally, following the Bloch's theorem the total Hamiltonian of the zigzag nanoribbon can be expressed as
\begin{equation}\label{ribb}
(E \mathcal{I}-\mathcal{E})=\mathcal{T} e^{i k_x a}+\mathcal{T}^{\dagger} e^{-i k_x a}\ .
\end{equation}
where $a$ is the unit cell separation. Here, $\mathcal{T}$ and $\mathcal{T}^{\dagger}$ describe the coupling of a unit cell
with the right and left adjacent supercells, respectively. The above equation can be solved numerically to obtain the
energy dispersion of the nanoribbon. 

\begin{figure}[!thpb]
\centering
\includegraphics[height=3.0cm,width=0.45\linewidth]{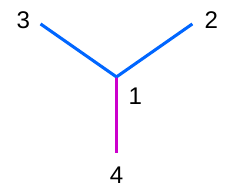}
\caption{Schematic of the three nearest neighbors, denoted by $2$, $3$ and $4$, are connected to $1$.}
\label{atoms}
\end{figure}
\begin{figure*}
 {\includegraphics[width=.32\textwidth,height=6cm]{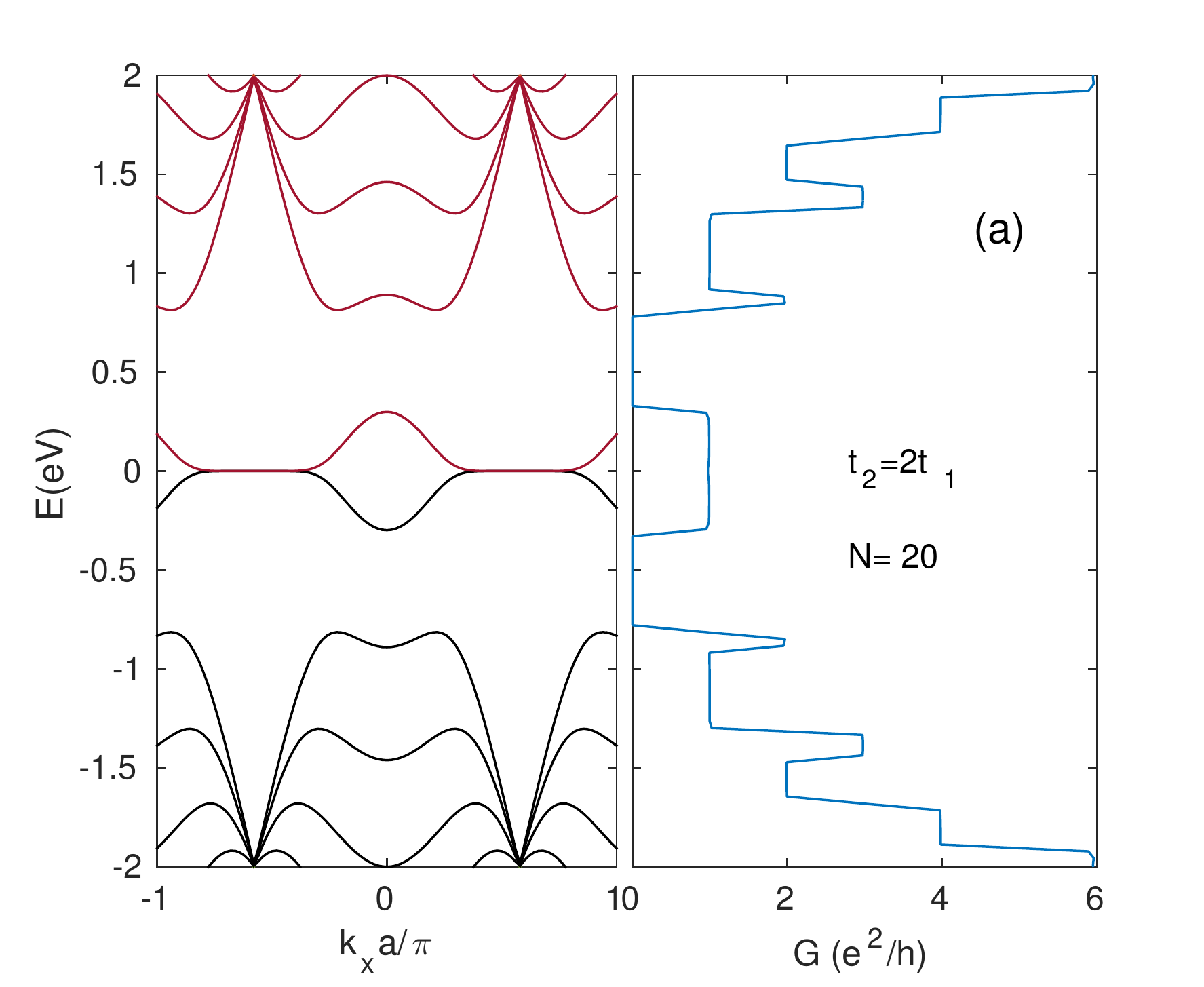}}
 {\includegraphics[width=.32\textwidth,height=6cm]{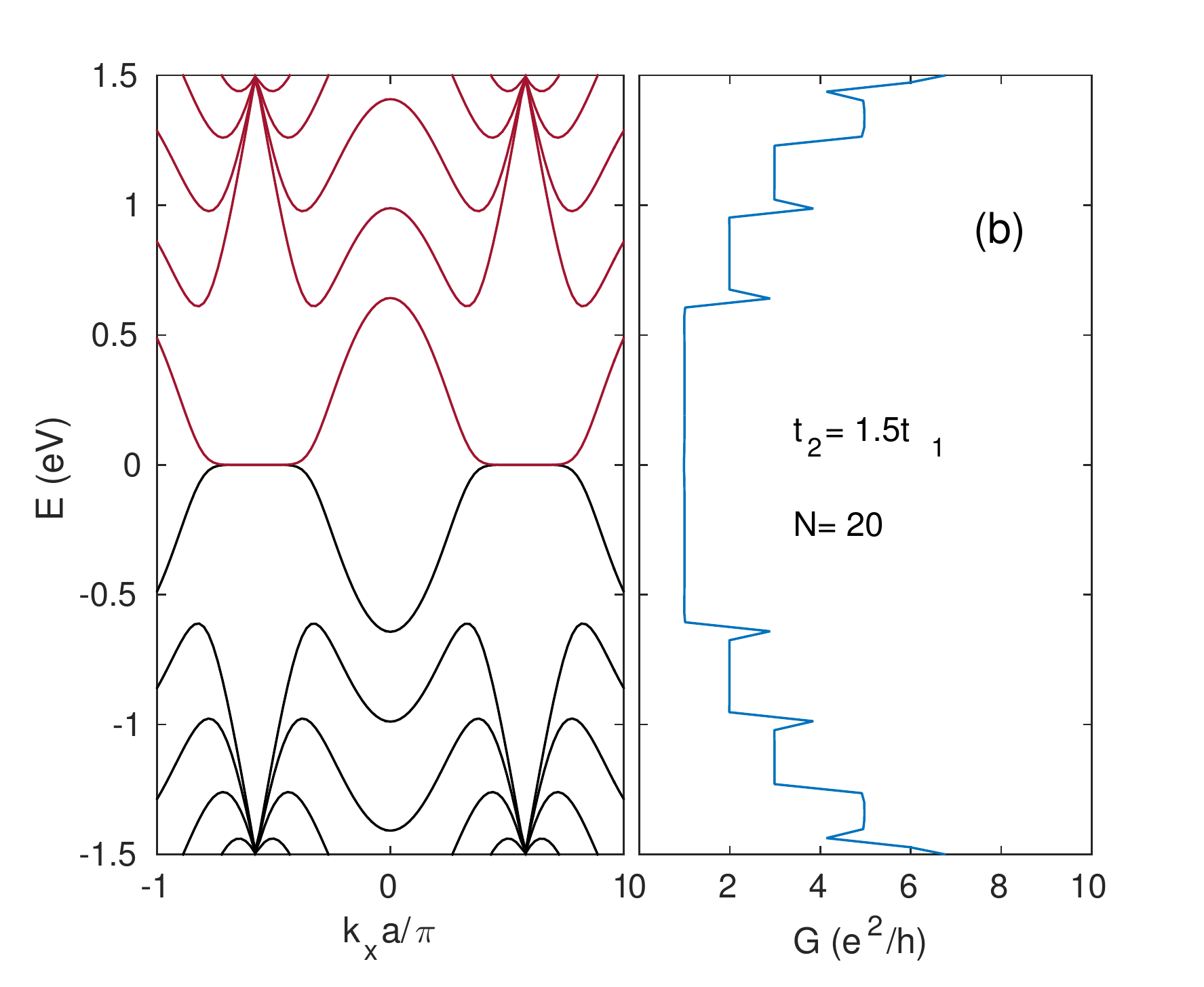}}
 {\includegraphics[width=.32\textwidth,height=6cm]{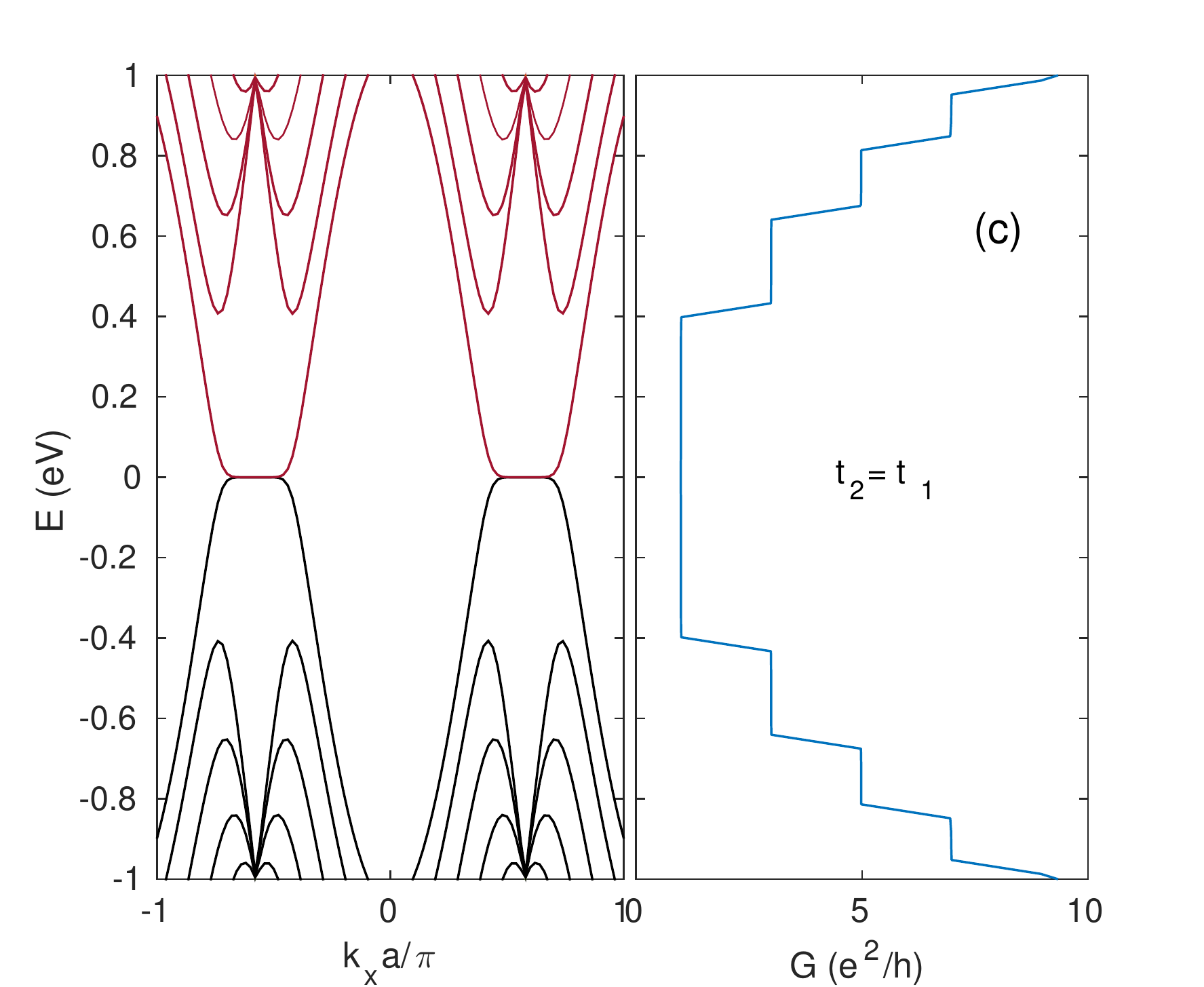}}
\caption{(Color online) Band dispersion and conductance are illustrated for there cases (a) $t_2=2t_1$ (semi-Dirac)
(b) $t_2=1.5t_1$ (c) $t_2=t_1$ (graphene).}
\label{cond_plot}
\end{figure*}

The energy band dispersion of the ZNR, based on the semi-Dirac material, is shown in Fig.~\ref{ribbo_con}.
This indicates that the band dispersion exhibits a pair of edge modes which are fully separated from the bulk. 
This is in contrast to the case of ZNR based on monolayer graphene~\cite{RevModPhys.81.109}
or silicene~\cite{PhysRevB.92.035413} where the edge modes are coupled to the bulk at the two valleys. 
Moreover, unlike the graphene, ZNR of the semi-Dirac material possesses a sufficient gap ($>1.5t_1$) which can be 
attributed to the presence of unequal hopping parameters in the lattice structure.

Note that, the appearance of edge modes in Fig.~\ref{ribbo_con} exhibit quite similar nature 
as in the case of graphene nanoribbon. In both the materials, semi-Dirac and graphene, the 
edge modes are strongly dependent on the edge geometry/commensurability of the lattice. Apart from this,
in both the materials the bulk is always gapless and these edge modes are not protected against the presence of
disorder etc. Hence, they don't have a topological character.

Moreover, we would like to reveal how the band structure as well as the conductance regain the 
properties of graphene with the variation of $t_2$ from $2t_1$ to $t_1$ via $1.5t_1$. The corresponding 
conductances are evaluated numerically by using standard recursive Green's function formalism~\cite{dutta,PhysRevB.92.035413}
which is given by
\begin{equation}
 G=\frac{e^2}{h}{\rm Tr}[\Gamma_L(E)G_{D}^{\dagger}\Gamma_R(E)G_{D}(E)]\ ,
\end{equation}
where $\Gamma_{L/R}(E)=i[\Sigma_{L/R}(E)-\Sigma^{\dagger}_{L/R}(E)]$ with $\Sigma_{L/R}(E)$ is the self energy of the 
left (right) lead. Hence,
\begin{equation}
 G_{D}=1/[E-\mathcal{E}-\Sigma_{L}-\Sigma_{R}]\ .
\end{equation}
Here, $\mathcal{E}$ is the on-site Hamiltonian matrix (see Eq.(\ref{HNR})).
The band dispersion and corresponding conductances are demonstrated in Fig.~\ref{cond_plot}
for three different cases. This exhibits smooth transmutation from semi-Dirac to graphene via $t_2=1.5t_1$.
It also highlights how two valleys evolve from single valley \ie~from semi-Dirac to graphene and how conductance
plateaus change from $t_{2}=2t_{1}$ to $t_{2}=t_{1}$ case. Note that the conductance (in units of $e^2/h$) 
corresponding to the edge modes in semi-Dirac material is clearly separated from the bulk conductance as
shown in Fig.~\ref{cond_plot}(a). This feature confirms the complete isolation of the edge modes from the bulk.
Another noticeable point here is that the conductane steps are not increasing
in ascending order due to the peculier nature of the band structure. On the other hand, in graphene, it increases
regular stepwise corresponding to each transverse mode (see Fig.~\ref{cond_plot}(c)).

\subsection{Irradiated ribbon}
\begin{figure*}[!thpb]
 {\includegraphics[width=.49\textwidth,height=6cm]{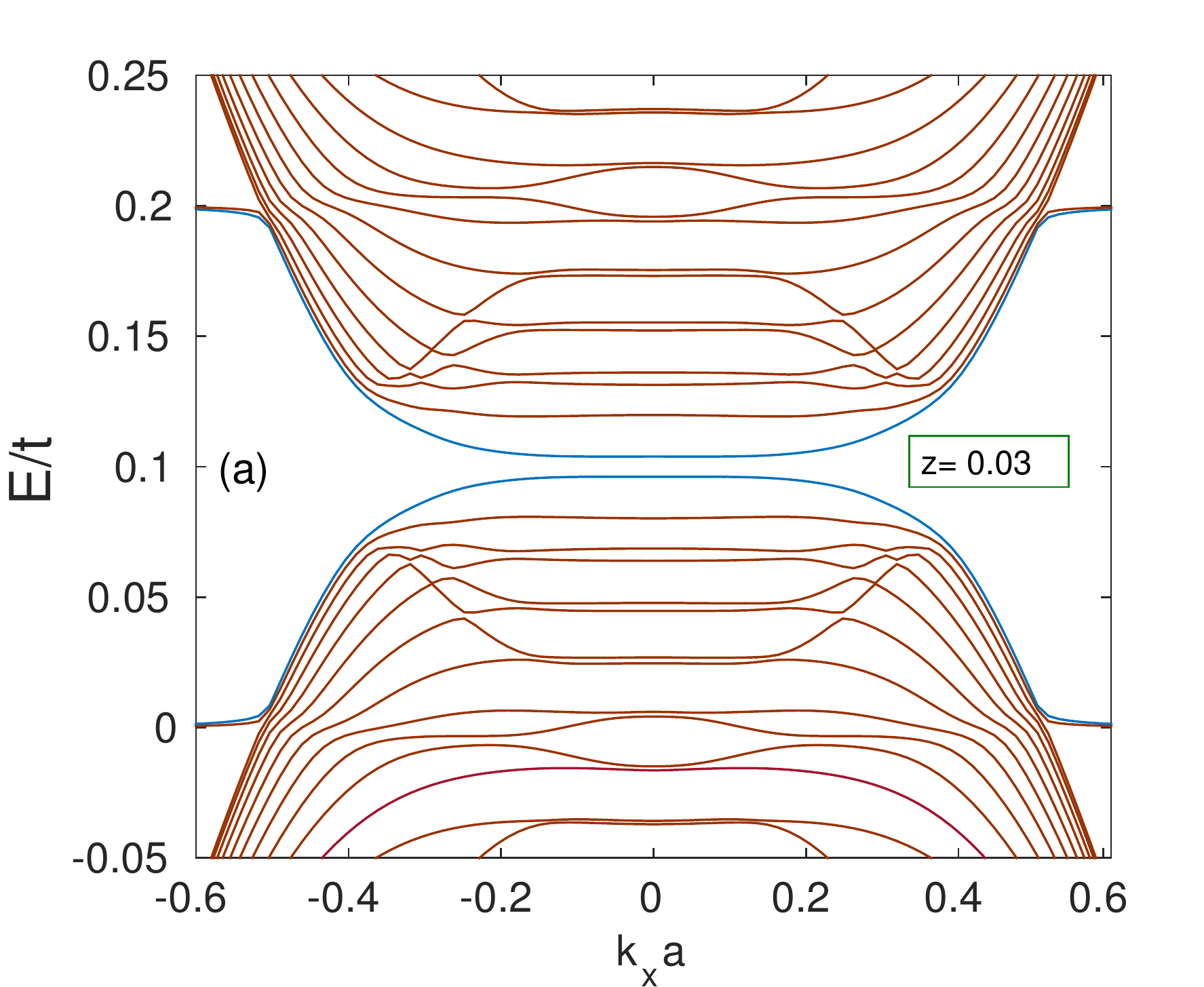}}
 {\includegraphics[width=.49\textwidth,height=6cm]{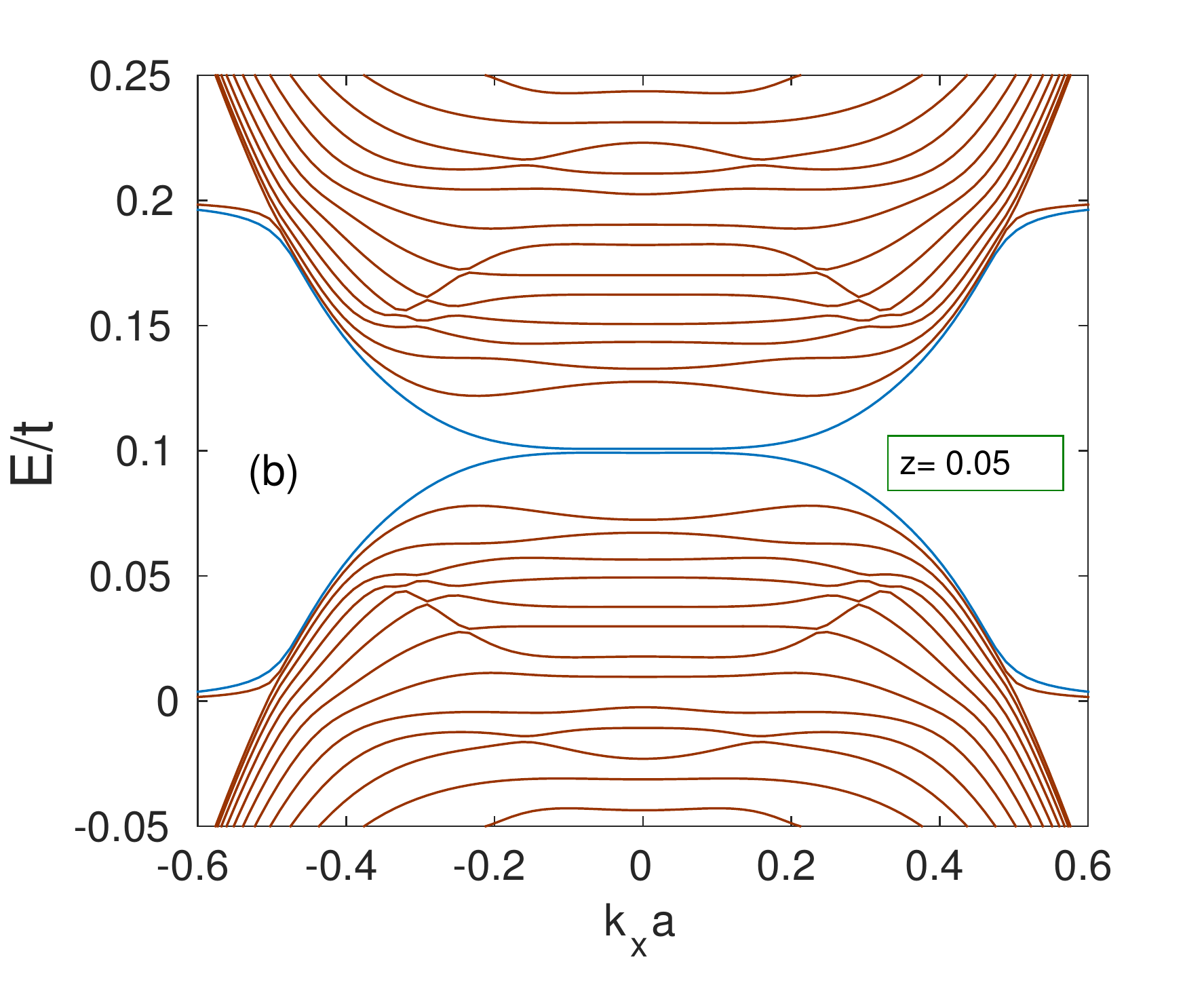}}
\caption{Quasi energy band dispersion of the irradiated semi-Dirac ZNR geometry is illustrated for (a) $z=0.03$ (b) $z=0.05$. Here $z=2\pi a_{cc}A_0/\phi_0$ 
describes the strength of the external periodic driving. The width of each supercell is given by $N=600$ and $\hbar\omega=0.2t_1$.}
\label{floq_ribbo_band}
\end{figure*}

In this subsection, we include the effect of irradiation in the band structure of the semi-Dirac material based ZNR.
To obtain the band dispersion we diagonalize the following Floquet tight-binding Hamiltonian as
\begin{equation}\label{floq_ribb}
 H_{F}=\left[\left(\mathcal{\tilde{E}}+\tilde{\mathcal{T}}_{F}e^{i k_x a}+
 \tilde{\mathcal{T}}_{F}^{\dagger} e^{-i k_x a}\right)_{N\times N}-n\hbar\omega\mathcal{I}
 \delta_{n,n'}\right]_{N\times n}\ .
\end{equation}
in the Floquet extended space. The size of the Floquet Hamiltonian is $N\times n$ where $n$ denotes the 
Floque replicas (side bands), representing different number of photons. The above Hamiltonian is a 
block matrix where each block is of $N\times N$ dimension.
The Floquet on-site energy is now denoted by $\mathcal{\tilde{E}}$.
The effect of irradiation on the band structure can be included by modifying 
the hoping parameters between $i^{\rm th}$ and $j^{\rm th}$ sublattice as
\begin{equation}
 t_{ij}=t_{ij}\exp\left[i\frac{2\pi}{\phi_0}\int_{r_i}^{r_j}A(t).dr\right]\ .
\end{equation}
After integrating out the time part, the three nearest neighbor hopping parameters that are left, can be written as 
\begin{equation}
 t_{12}=t_{1}e^{im\pi/6}\mathcal{J}_{-m}(-z)\ ,
\end{equation}
\begin{equation}
 t_{13}=t_{1}e^{i5m\pi/6}\mathcal{J}_{-m}(z)\ ,
\end{equation}
and 
\begin{equation}
 t_{14}=t_{2}e^{-im\pi/2}\mathcal{J}_{-m}(-z) \ .
\end{equation}
Here, $\mathcal{J}_{m}(z)$ represents the Bessel function of the first kind of order $m (=n-n')$.
The schematic of the three nearest neighbor hopping parameters, between $i$ and $j$ with $i=1$ 
and $j=2,3,4$, is shown in Fig.~\ref{atoms}. The above mentioned emergent hopping parameters are taken 
into account in Eq.(\ref{floq_ribb}) in order to numerically diagonalize it in the extended Floquet space.
The quasi energy band dispersion is obtained numerically and depicted in Fig.~\ref{floq_ribbo_band}.
Here, the dimensionless parameter $z=2\pi A_0 a/\phi_0$ denotes the strength of periodic driving with $\phi_0$
being the magnetic flux quanta. In our analysis, the number of Floquet replicas (side bands) are 
considered to be $n=2$. This is sufficient to capture the relevant features of the band structure.
Any further enhancement of the number of side bands only assembles more quasi energy modes 
without producing any new qualitative features to the band structure. Unlike the case of non-irradiated band structure 
of ZNR (see Fig~\ref{ribbo_con}), the edge modes (sky blue lines) in the irradiated case are not fully separated from the bulk. 
Also the dynamical band gap, at $\hbar\omega/2$, appears to be vanishingly small (see Fig.~\ref{floq_ribbo_band}(a)).
Moreover, the band gap can be further reduced to be almost zero if we increase the amplitude of the irradiation,
as shown in Fig.~\ref{floq_ribbo_band}(b). Hence, one can infer that by applying external periodic driving
the decoupled edge modes can be enforced to couple the bulk in semi-Dirac material. Note that, in graphene
nanoribbon, the  irradiation (circularly polarized light) can give rise to topological chiral Floquet edge modes~\cite{PhysRevB.89.121401} 
around the dynamical gap at $\hbar\omega/2$. However, such feature is absent in case of semi-Dirac ZNR. 
The reason can be attributed to the fact that circularly polarized light can open up
a gap into the bulk of graphene, but not in semi-Dirac material ($k=0$ point is always gapless).
Hence, Floquet edge modes in semi-Dirac material do not seem to have a topological character.
Nevertheless, we also check that by setting $t_2=t_1$, the chiral Floquet edge modes of graphene 
can be recovered similar to the non-irradiated case.

\begin{figure}[!thpb]
\centering
\includegraphics[height=6cm,width=\linewidth]{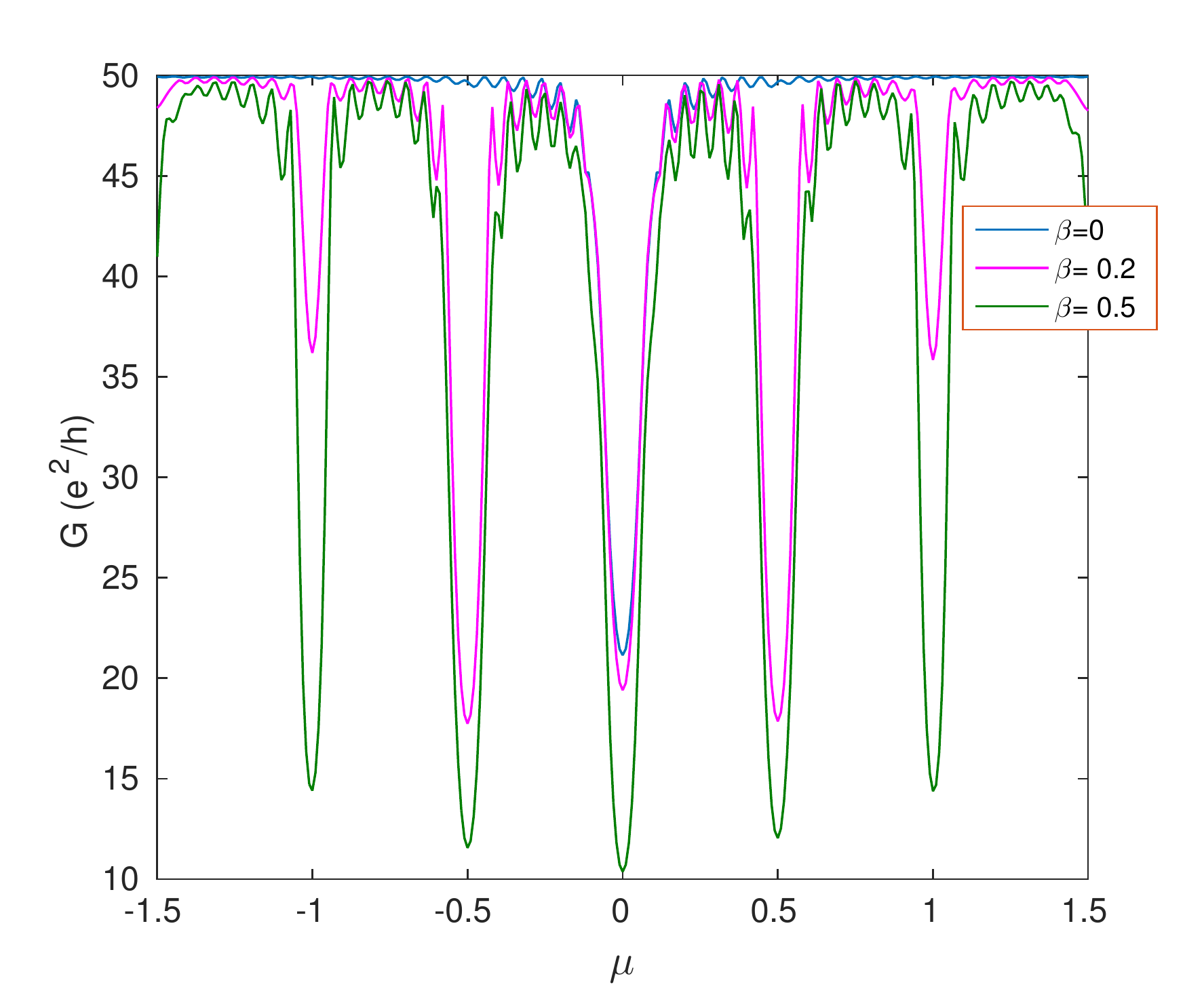}
\caption{(Color online) The conductance spectrum of the ZNR geometry is demonstrated as a function of the chemical
potential $\mu$ for various strengths of irradiation.}
\label{floq_ribbon_con}
\end{figure}

We also explore the transport signature of the ZNR geometry. To obtain the conductance,
we extend the formalism used for the bulk by taking into account the quantized transverse wave vector,
determined by the width of the ribbon as $k_y=n\pi/W$ with $n=1,2,3..$ being the number of transverse edge modes.
Note that, in this method the edge modes cannot be captured, rather mostly transverse modes reside inside the bulk.
This approximate method is already implemented in case of graphene~\cite{PhysRevB.96.245404}.
In our case, as the edge modes are not chiral in nature, we use the same formalism as used for 
the bulk instead of the formalism described in Refs.~[\onlinecite{kohler2005driven,PhysRevB.72.245339}].
The conductance for the ZNR is shown in Fig.~\ref{floq_ribbon_con}. The behavior of the conductance
exhibits the dips at $\mu=0$ and $\pm \hbar\omega/2$. This is quite similar to the case of the bulk semi-Dirac material.
However, the conductance varies very slowly with chemical potential except at the locations of the dips. 
In our numerical analysis, we consider the number of transverse edge modes to be $n=50$ for which conductance 
remains weakly oscillatory around $50 e^2/h$. A more careful analysis of the conductance for semi-Dirac ZNR 
geometry, based on Floquet-Landauer Green's function approach~\cite{kohler2005driven,PhysRevB.72.245339} 
is beyond the scope of the present work and will be presented elsewhere. 

\section{Summary and conclusions}\label{sec5}
In our work, we investigate the electronic and transport properties of an irradiated semi-Dirac material. 
We consider the irradiation in the form of a circularly polarized light. We present our results for the semi-Dirac material 
in comparison to the irradiated monolayer graphene. The irradiation is unable to open up a gap in the Floquet spectrum 
of semi-Dirac material within the high frequency approximation~\cite{PhysRevB.91.205445}. 
However, beyond this approximation, a full band dispersion analysis reveals that the band gap indeed appears 
for higher values of the momentum and between two nearest Floquet side bands. Such appearance of the
band gap in the quasi energy spectrum can also be probed via the signature of the conductance spectrum. 
The latter manifests several dips around the gaps. We also observe that the band gap opening is strongly dependent 
on the angular orientation of momentum due to the anisotropic band structure of the semi-Dirac material. 
We also explore the band structure of nanoribbon geometry comprised of such material and show some 
distinct features in comparison to the manolayer graphene nanoribbon.
The ZNR of the semi-Dirac material hosts a pair of edge modes which are fully detached from the bulk. 
Also, they smoothly transmute to the graphene edge modes when the two hopping parameters $t_{1}$ and $t_{2}$ become equal.
Furthermore, in presence of an external irradiation, ZNR of such material does not host topologically protected 
Floquet chiral edge modes like graphene. Moreover, the edge modes become coupled to the bulk and the dynamical band gap 
in the Floquet spectrum begins to disappear with the increase of the strength of the periodic drive. 
Finally, we also explore the possible features of the conductance of the irradiated ZNR based on semi-Dirac material. 
The conductance exhibits similar dips around the same locations of the chemical potential as in the bulk except the 
amplitude mismatch due to the presence of finite number of channels.

Finally, we discuss the experimental feasibility of transport measurement in an 
externally irradiated semi-Dirac material. As the transport (current) measurement of an irradiated topological insulator 
thin film has been successfully performed~\cite{zhang2014anomalous}, we expect that the conductance
dips may also be observed in semi-Dirac material via the two-terminal measurement. In practical situation, the chemical potential 
(carrier density) in Dirac/semi-Dirac material can be tuned by applying external gate voltage in the non-irradiated region. 
Then, if one is able to measure the two-terminal conductance of our set-up, then the dips that appear in our angle averaged conductance spectrum can be an indirect signature of the gapped region. However, in experiment, one may have to encounter the presence of disorder or a small deviation from the perfect circularly polarized light. Nevertheless, it has been recently reported in another hexagonal 2D lattice that disorder causes an overall suppression to the conductance~\cite{tahir2016floquet} and it cannot destroy the conductance dips. On the other hand, the elliptical polarized light does not cause significant changes in the results of circularly polarized light~\cite{zhang2014anomalous,tahir2016floquet}.

\begin{acknowledgements}
SFI acknowledges Jonathan Atteia, P. M. Perez-Piskunow and Paramita Dutta for useful and stimulating discussions.
\end{acknowledgements}
\bibliography{bibfile}{}

\begin{thebibliography}{39}%
\makeatletter
\providecommand \@ifxundefined [1]{%
 \@ifx{#1\undefined}
}%
\providecommand \@ifnum [1]{%
 \ifnum #1\expandafter \@firstoftwo
 \else \expandafter \@secondoftwo
 \fi
}%
\providecommand \@ifx [1]{%
 \ifx #1\expandafter \@firstoftwo
 \else \expandafter \@secondoftwo
 \fi
}%
\providecommand \natexlab [1]{#1}%
\providecommand \enquote  [1]{``#1''}%
\providecommand \bibnamefont  [1]{#1}%
\providecommand \bibfnamefont [1]{#1}%
\providecommand \citenamefont [1]{#1}%
\providecommand \href@noop [0]{\@secondoftwo}%
\providecommand \href [0]{\begingroup \@sanitize@url \@href}%
\providecommand \@href[1]{\@@startlink{#1}\@@href}%
\providecommand \@@href[1]{\endgroup#1\@@endlink}%
\providecommand \@sanitize@url [0]{\catcode `\\12\catcode `\$12\catcode
  `\&12\catcode `\#12\catcode `\^12\catcode `\_12\catcode `\%12\relax}%
\providecommand \@@startlink[1]{}%
\providecommand \@@endlink[0]{}%
\providecommand \url  [0]{\begingroup\@sanitize@url \@url }%
\providecommand \@url [1]{\endgroup\@href {#1}{\urlprefix }}%
\providecommand \urlprefix  [0]{URL }%
\providecommand \Eprint [0]{\href }%
\providecommand \doibase [0]{http://dx.doi.org/}%
\providecommand \selectlanguage [0]{\@gobble}%
\providecommand \bibinfo  [0]{\@secondoftwo}%
\providecommand \bibfield  [0]{\@secondoftwo}%
\providecommand \translation [1]{[#1]}%
\providecommand \BibitemOpen [0]{}%
\providecommand \bibitemStop [0]{}%
\providecommand \bibitemNoStop [0]{.\EOS\space}%
\providecommand \EOS [0]{\spacefactor3000\relax}%
\providecommand \BibitemShut  [1]{\csname bibitem#1\endcsname}%
\let\auto@bib@innerbib\@empty
\bibitem [{\citenamefont {Oka}\ and\ \citenamefont
  {Aoki}(2009)}]{PhysRevB.79.081406}%
  \BibitemOpen
  \bibfield  {author} {\bibinfo {author} {\bibfnamefont {T.}~\bibnamefont
  {Oka}}\ and\ \bibinfo {author} {\bibfnamefont {H.}~\bibnamefont {Aoki}},\
  }\href@noop {} {\bibfield  {journal} {\bibinfo  {journal} {Phys. Rev. B}\
  }\textbf {\bibinfo {volume} {79}},\ \bibinfo {pages} {081406} (\bibinfo
  {year} {2009})}\BibitemShut {NoStop}%
\bibitem [{\citenamefont {Lindner}\ \emph {et~al.}(2011)\citenamefont
  {Lindner}, \citenamefont {Refael},\ and\ \citenamefont
  {Galitski}}]{lindner2011floquet}%
  \BibitemOpen
  \bibfield  {author} {\bibinfo {author} {\bibfnamefont {N.~H.}\ \bibnamefont
  {Lindner}}, \bibinfo {author} {\bibfnamefont {G.}~\bibnamefont {Refael}}, \
  and\ \bibinfo {author} {\bibfnamefont {V.}~\bibnamefont {Galitski}},\
  }\href@noop {} {\bibfield  {journal} {\bibinfo  {journal} {Nature Physics}\
  }\textbf {\bibinfo {volume} {7}},\ \bibinfo {pages} {490} (\bibinfo {year}
  {2011})}\BibitemShut {NoStop}%
\bibitem [{\citenamefont {Inoue}\ and\ \citenamefont
  {Tanaka}(2012)}]{PhysRevB.85.125425}%
  \BibitemOpen
  \bibfield  {author} {\bibinfo {author} {\bibfnamefont {J.-i.}\ \bibnamefont
  {Inoue}}\ and\ \bibinfo {author} {\bibfnamefont {A.}~\bibnamefont {Tanaka}},\
  }\href@noop {} {\bibfield  {journal} {\bibinfo  {journal} {Phys. Rev. B}\
  }\textbf {\bibinfo {volume} {85}},\ \bibinfo {pages} {125425} (\bibinfo
  {year} {2012})}\BibitemShut {NoStop}%
\bibitem [{\citenamefont {Kitagawa}\ \emph {et~al.}(2011)\citenamefont
  {Kitagawa}, \citenamefont {Oka}, \citenamefont {Brataas}, \citenamefont
  {Fu},\ and\ \citenamefont {Demler}}]{PhysRevB.84.235108}%
  \BibitemOpen
  \bibfield  {author} {\bibinfo {author} {\bibfnamefont {T.}~\bibnamefont
  {Kitagawa}}, \bibinfo {author} {\bibfnamefont {T.}~\bibnamefont {Oka}},
  \bibinfo {author} {\bibfnamefont {A.}~\bibnamefont {Brataas}}, \bibinfo
  {author} {\bibfnamefont {L.}~\bibnamefont {Fu}}, \ and\ \bibinfo {author}
  {\bibfnamefont {E.}~\bibnamefont {Demler}},\ }\href@noop {} {\bibfield
  {journal} {\bibinfo  {journal} {Phys. Rev. B}\ }\textbf {\bibinfo {volume}
  {84}},\ \bibinfo {pages} {235108} (\bibinfo {year} {2011})}\BibitemShut
  {NoStop}%
\bibitem [{\citenamefont {L\'opez}\ \emph {et~al.}(2012)\citenamefont
  {L\'opez}, \citenamefont {Sun},\ and\ \citenamefont
  {Schliemann}}]{PhysRevB.85.205428}%
  \BibitemOpen
  \bibfield  {author} {\bibinfo {author} {\bibfnamefont {A.}~\bibnamefont
  {L\'opez}}, \bibinfo {author} {\bibfnamefont {Z.~Z.}\ \bibnamefont {Sun}}, \
  and\ \bibinfo {author} {\bibfnamefont {J.}~\bibnamefont {Schliemann}},\
  }\href@noop {} {\bibfield  {journal} {\bibinfo  {journal} {Phys. Rev. B}\
  }\textbf {\bibinfo {volume} {85}},\ \bibinfo {pages} {205428} (\bibinfo
  {year} {2012})}\BibitemShut {NoStop}%
\bibitem [{\citenamefont {Zhai}\ and\ \citenamefont
  {Jin}(2014)}]{PhysRevB.89.235416}%
  \BibitemOpen
  \bibfield  {author} {\bibinfo {author} {\bibfnamefont {X.}~\bibnamefont
  {Zhai}}\ and\ \bibinfo {author} {\bibfnamefont {G.}~\bibnamefont {Jin}},\
  }\href@noop {} {\bibfield  {journal} {\bibinfo  {journal} {Phys. Rev. B}\
  }\textbf {\bibinfo {volume} {89}},\ \bibinfo {pages} {235416} (\bibinfo
  {year} {2014})}\BibitemShut {NoStop}%
\bibitem [{\citenamefont {Saha}(2016)}]{PhysRevB.94.081103}%
  \BibitemOpen
  \bibfield  {author} {\bibinfo {author} {\bibfnamefont {K.}~\bibnamefont
  {Saha}},\ }\href@noop {} {\bibfield  {journal} {\bibinfo  {journal} {Phys.
  Rev. B}\ }\textbf {\bibinfo {volume} {94}},\ \bibinfo {pages} {081103}
  (\bibinfo {year} {2016})}\BibitemShut {NoStop}%
\bibitem [{\citenamefont {Ezawa}(2013)}]{ezawa2013photoinduced}%
  \BibitemOpen
  \bibfield  {author} {\bibinfo {author} {\bibfnamefont {M.}~\bibnamefont
  {Ezawa}},\ }\href@noop {} {\bibfield  {journal} {\bibinfo  {journal} {Phys.
  Rev. Lett.}\ }\textbf {\bibinfo {volume} {110}},\ \bibinfo {pages} {026603}
  (\bibinfo {year} {2013})}\BibitemShut {NoStop}%
\bibitem [{\citenamefont {Perez-Piskunow}\ \emph {et~al.}(2014)\citenamefont
  {Perez-Piskunow}, \citenamefont {Usaj}, \citenamefont {Balseiro},\ and\
  \citenamefont {Torres}}]{PhysRevB.89.121401}%
  \BibitemOpen
  \bibfield  {author} {\bibinfo {author} {\bibfnamefont {P.~M.}\ \bibnamefont
  {Perez-Piskunow}}, \bibinfo {author} {\bibfnamefont {G.}~\bibnamefont
  {Usaj}}, \bibinfo {author} {\bibfnamefont {C.~A.}\ \bibnamefont {Balseiro}},
  \ and\ \bibinfo {author} {\bibfnamefont {L.~E. F.~F.}\ \bibnamefont
  {Torres}},\ }\href@noop {} {\bibfield  {journal} {\bibinfo  {journal} {Phys.
  Rev. B}\ }\textbf {\bibinfo {volume} {89}},\ \bibinfo {pages} {121401}
  (\bibinfo {year} {2014})}\BibitemShut {NoStop}%
\bibitem [{\citenamefont {Cayssol}\ \emph {et~al.}(2013)\citenamefont
  {Cayssol}, \citenamefont {D{\'o}ra}, \citenamefont {Simon},\ and\
  \citenamefont {Moessner}}]{cayssol2013floquet}%
  \BibitemOpen
  \bibfield  {author} {\bibinfo {author} {\bibfnamefont {J.}~\bibnamefont
  {Cayssol}}, \bibinfo {author} {\bibfnamefont {B.}~\bibnamefont {D{\'o}ra}},
  \bibinfo {author} {\bibfnamefont {F.}~\bibnamefont {Simon}}, \ and\ \bibinfo
  {author} {\bibfnamefont {R.}~\bibnamefont {Moessner}},\ }\href@noop {}
  {\bibfield  {journal} {\bibinfo  {journal} {Physica Status Solidi
  (RRL)--Rapid Research Letters}\ }\textbf {\bibinfo {volume} {7}},\ \bibinfo
  {pages} {101} (\bibinfo {year} {2013})}\BibitemShut {NoStop}%
\bibitem [{\citenamefont {Peng}\ \emph {et~al.}(2016)\citenamefont {Peng},
  \citenamefont {Qin}, \citenamefont {Zhao}, \citenamefont {Shen},
  \citenamefont {Xu}, \citenamefont {Bao}, \citenamefont {Jia},\ and\
  \citenamefont {Zhu}}]{peng2016experimental}%
  \BibitemOpen
  \bibfield  {author} {\bibinfo {author} {\bibfnamefont {Y.-G.}\ \bibnamefont
  {Peng}}, \bibinfo {author} {\bibfnamefont {C.-Z.}\ \bibnamefont {Qin}},
  \bibinfo {author} {\bibfnamefont {D.-G.}\ \bibnamefont {Zhao}}, \bibinfo
  {author} {\bibfnamefont {Y.-X.}\ \bibnamefont {Shen}}, \bibinfo {author}
  {\bibfnamefont {X.-Y.}\ \bibnamefont {Xu}}, \bibinfo {author} {\bibfnamefont
  {M.}~\bibnamefont {Bao}}, \bibinfo {author} {\bibfnamefont {H.}~\bibnamefont
  {Jia}}, \ and\ \bibinfo {author} {\bibfnamefont {X.-F.}\ \bibnamefont
  {Zhu}},\ }\href@noop {} {\bibfield  {journal} {\bibinfo  {journal} {Nature
  Communications}\ }\textbf {\bibinfo {volume} {7}},\ \bibinfo {pages} {13368}
  (\bibinfo {year} {2016})}\BibitemShut {NoStop}%
\bibitem [{\citenamefont {Zhang}\ \emph {et~al.}(2014)\citenamefont {Zhang},
  \citenamefont {Yao}, \citenamefont {Shao}, \citenamefont {Li}, \citenamefont
  {Li}, \citenamefont {Bao}, \citenamefont {Wang},\ and\ \citenamefont
  {Yang}}]{zhang2014anomalous}%
  \BibitemOpen
  \bibfield  {author} {\bibinfo {author} {\bibfnamefont {H.}~\bibnamefont
  {Zhang}}, \bibinfo {author} {\bibfnamefont {J.}~\bibnamefont {Yao}}, \bibinfo
  {author} {\bibfnamefont {J.}~\bibnamefont {Shao}}, \bibinfo {author}
  {\bibfnamefont {H.}~\bibnamefont {Li}}, \bibinfo {author} {\bibfnamefont
  {S.}~\bibnamefont {Li}}, \bibinfo {author} {\bibfnamefont {D.}~\bibnamefont
  {Bao}}, \bibinfo {author} {\bibfnamefont {C.}~\bibnamefont {Wang}}, \ and\
  \bibinfo {author} {\bibfnamefont {G.}~\bibnamefont {Yang}},\ }\href@noop {}
  {\bibfield  {journal} {\bibinfo  {journal} {Scientific reports}\ }\textbf
  {\bibinfo {volume} {4}},\ \bibinfo {pages} {5876} (\bibinfo {year}
  {2014})}\BibitemShut {NoStop}%
\bibitem [{\citenamefont {Wang}\ \emph {et~al.}(2013)\citenamefont {Wang},
  \citenamefont {Steinberg}, \citenamefont {Jarillo-Herrero},\ and\
  \citenamefont {Gedik}}]{wang2013observation}%
  \BibitemOpen
  \bibfield  {author} {\bibinfo {author} {\bibfnamefont {Y.}~\bibnamefont
  {Wang}}, \bibinfo {author} {\bibfnamefont {H.}~\bibnamefont {Steinberg}},
  \bibinfo {author} {\bibfnamefont {P.}~\bibnamefont {Jarillo-Herrero}}, \ and\
  \bibinfo {author} {\bibfnamefont {N.}~\bibnamefont {Gedik}},\ }\href@noop {}
  {\bibfield  {journal} {\bibinfo  {journal} {Science}\ }\textbf {\bibinfo
  {volume} {342}},\ \bibinfo {pages} {453} (\bibinfo {year}
  {2013})}\BibitemShut {NoStop}%
\bibitem [{\citenamefont {Golub}\ \emph {et~al.}(2011)\citenamefont {Golub},
  \citenamefont {Tarasenko}, \citenamefont {Entin},\ and\ \citenamefont
  {Magarill}}]{PhysRevB.84.195408}%
  \BibitemOpen
  \bibfield  {author} {\bibinfo {author} {\bibfnamefont {L.~E.}\ \bibnamefont
  {Golub}}, \bibinfo {author} {\bibfnamefont {S.~A.}\ \bibnamefont
  {Tarasenko}}, \bibinfo {author} {\bibfnamefont {M.~V.}\ \bibnamefont
  {Entin}}, \ and\ \bibinfo {author} {\bibfnamefont {L.~I.}\ \bibnamefont
  {Magarill}},\ }\href@noop {} {\bibfield  {journal} {\bibinfo  {journal}
  {Phys. Rev. B}\ }\textbf {\bibinfo {volume} {84}},\ \bibinfo {pages} {195408}
  (\bibinfo {year} {2011})}\BibitemShut {NoStop}%
\bibitem [{\citenamefont {Dey}\ and\ \citenamefont {Ghosh}(2018)}]{tarun}%
  \BibitemOpen
  \bibfield  {author} {\bibinfo {author} {\bibfnamefont {B.}~\bibnamefont
  {Dey}}\ and\ \bibinfo {author} {\bibfnamefont {T.~K.}\ \bibnamefont
  {Ghosh}},\ }\href@noop {} {\bibfield  {journal} {\bibinfo  {journal} {Phys.
  Rev. B}\ }\textbf {\bibinfo {volume} {98}},\ \bibinfo {pages} {075422}
  (\bibinfo {year} {2018})}\BibitemShut {NoStop}%
\bibitem [{\citenamefont {Tahir}\ \emph {et~al.}(2014)\citenamefont {Tahir},
  \citenamefont {Manchon},\ and\ \citenamefont
  {Schwingenschl\"ogl}}]{PhysRevB.90.125438}%
  \BibitemOpen
  \bibfield  {author} {\bibinfo {author} {\bibfnamefont {M.}~\bibnamefont
  {Tahir}}, \bibinfo {author} {\bibfnamefont {A.}~\bibnamefont {Manchon}}, \
  and\ \bibinfo {author} {\bibfnamefont {U.}~\bibnamefont
  {Schwingenschl\"ogl}},\ }\href@noop {} {\bibfield  {journal} {\bibinfo
  {journal} {Phys. Rev. B}\ }\textbf {\bibinfo {volume} {90}},\ \bibinfo
  {pages} {125438} (\bibinfo {year} {2014})}\BibitemShut {NoStop}%
\bibitem [{\citenamefont {Tahir}\ \emph {et~al.}(2016)\citenamefont {Tahir},
  \citenamefont {Zhang},\ and\ \citenamefont
  {Schwingenschl{\"o}gl}}]{tahir2016floquet}%
  \BibitemOpen
  \bibfield  {author} {\bibinfo {author} {\bibfnamefont {M.}~\bibnamefont
  {Tahir}}, \bibinfo {author} {\bibfnamefont {Q.}~\bibnamefont {Zhang}}, \ and\
  \bibinfo {author} {\bibfnamefont {U.}~\bibnamefont {Schwingenschl{\"o}gl}},\
  }\href@noop {} {\bibfield  {journal} {\bibinfo  {journal} {Scientific
  reports}\ }\textbf {\bibinfo {volume} {6}},\ \bibinfo {pages} {31821}
  (\bibinfo {year} {2016})}\BibitemShut {NoStop}%
\bibitem [{\citenamefont {Tahir}\ and\ \citenamefont
  {Schwingenschl{\"o}gl}(2014)}]{tahir2014tunable}%
  \BibitemOpen
  \bibfield  {author} {\bibinfo {author} {\bibfnamefont {M.}~\bibnamefont
  {Tahir}}\ and\ \bibinfo {author} {\bibfnamefont {U.}~\bibnamefont
  {Schwingenschl{\"o}gl}},\ }\href@noop {} {\bibfield  {journal} {\bibinfo
  {journal} {New J. of Phys.}\ }\textbf {\bibinfo {volume} {16}},\ \bibinfo
  {pages} {115003} (\bibinfo {year} {2014})}\BibitemShut {NoStop}%
\bibitem [{\citenamefont {Tahir}\ and\ \citenamefont
  {Vasilopoulos}(2015)}]{PhysRevB.91.115311}%
  \BibitemOpen
  \bibfield  {author} {\bibinfo {author} {\bibfnamefont {M.}~\bibnamefont
  {Tahir}}\ and\ \bibinfo {author} {\bibfnamefont {P.}~\bibnamefont
  {Vasilopoulos}},\ }\href@noop {} {\bibfield  {journal} {\bibinfo  {journal}
  {Phys. Rev. B}\ }\textbf {\bibinfo {volume} {91}},\ \bibinfo {pages} {115311}
  (\bibinfo {year} {2015})}\BibitemShut {NoStop}%
\bibitem [{\citenamefont {Zhou}\ and\ \citenamefont
  {Jin}(2016)}]{PhysRevB.94.165436}%
  \BibitemOpen
  \bibfield  {author} {\bibinfo {author} {\bibfnamefont {X.}~\bibnamefont
  {Zhou}}\ and\ \bibinfo {author} {\bibfnamefont {G.}~\bibnamefont {Jin}},\
  }\href@noop {} {\bibfield  {journal} {\bibinfo  {journal} {Phys. Rev. B}\
  }\textbf {\bibinfo {volume} {94}},\ \bibinfo {pages} {165436} (\bibinfo
  {year} {2016})}\BibitemShut {NoStop}%
\bibitem [{\citenamefont {Khanna}\ \emph {et~al.}(2017)\citenamefont {Khanna},
  \citenamefont {Rao},\ and\ \citenamefont {Kundu}}]{PhysRevB.95.201115}%
  \BibitemOpen
  \bibfield  {author} {\bibinfo {author} {\bibfnamefont {U.}~\bibnamefont
  {Khanna}}, \bibinfo {author} {\bibfnamefont {S.}~\bibnamefont {Rao}}, \ and\
  \bibinfo {author} {\bibfnamefont {A.}~\bibnamefont {Kundu}},\ }\href@noop {}
  {\bibfield  {journal} {\bibinfo  {journal} {Phys. Rev. B}\ }\textbf {\bibinfo
  {volume} {95}},\ \bibinfo {pages} {201115} (\bibinfo {year}
  {2017})}\BibitemShut {NoStop}%
\bibitem [{\citenamefont {Narayan}(2015)}]{PhysRevB.91.205445}%
  \BibitemOpen
  \bibfield  {author} {\bibinfo {author} {\bibfnamefont {A.}~\bibnamefont
  {Narayan}},\ }\href@noop {} {\bibfield  {journal} {\bibinfo  {journal} {Phys.
  Rev. B}\ }\textbf {\bibinfo {volume} {91}},\ \bibinfo {pages} {205445}
  (\bibinfo {year} {2015})}\BibitemShut {NoStop}%
\bibitem [{\citenamefont {Chen}\ \emph {et~al.}(2018)\citenamefont {Chen},
  \citenamefont {Du},\ and\ \citenamefont {Fiete}}]{PhysRevB.97.035422}%
  \BibitemOpen
  \bibfield  {author} {\bibinfo {author} {\bibfnamefont {Q.}~\bibnamefont
  {Chen}}, \bibinfo {author} {\bibfnamefont {L.}~\bibnamefont {Du}}, \ and\
  \bibinfo {author} {\bibfnamefont {G.~A.}\ \bibnamefont {Fiete}},\ }\href@noop
  {} {\bibfield  {journal} {\bibinfo  {journal} {Phys. Rev. B}\ }\textbf
  {\bibinfo {volume} {97}},\ \bibinfo {pages} {035422} (\bibinfo {year}
  {2018})}\BibitemShut {NoStop}%
\bibitem [{\citenamefont {Banerjee}\ \emph {et~al.}(2009)\citenamefont
  {Banerjee}, \citenamefont {Singh}, \citenamefont {Pardo},\ and\ \citenamefont
  {Pickett}}]{banerjee2009tight}%
  \BibitemOpen
  \bibfield  {author} {\bibinfo {author} {\bibfnamefont {S.}~\bibnamefont
  {Banerjee}}, \bibinfo {author} {\bibfnamefont {R.}~\bibnamefont {Singh}},
  \bibinfo {author} {\bibfnamefont {V.}~\bibnamefont {Pardo}}, \ and\ \bibinfo
  {author} {\bibfnamefont {W.}~\bibnamefont {Pickett}},\ }\href@noop {}
  {\bibfield  {journal} {\bibinfo  {journal} {Phys. Rev. Lett.}\ }\textbf
  {\bibinfo {volume} {103}},\ \bibinfo {pages} {016402} (\bibinfo {year}
  {2009})}\BibitemShut {NoStop}%
\bibitem [{\citenamefont {Dietl}\ \emph {et~al.}(2008)\citenamefont {Dietl},
  \citenamefont {Pi{\'e}chon},\ and\ \citenamefont
  {Montambaux}}]{dietl2008new}%
  \BibitemOpen
  \bibfield  {author} {\bibinfo {author} {\bibfnamefont {P.}~\bibnamefont
  {Dietl}}, \bibinfo {author} {\bibfnamefont {F.}~\bibnamefont {Pi{\'e}chon}},
  \ and\ \bibinfo {author} {\bibfnamefont {G.}~\bibnamefont {Montambaux}},\
  }\href@noop {} {\bibfield  {journal} {\bibinfo  {journal} {Phys. Rev. Lett.}\
  }\textbf {\bibinfo {volume} {100}},\ \bibinfo {pages} {236405} (\bibinfo
  {year} {2008})}\BibitemShut {NoStop}%
\bibitem [{\citenamefont {Pyatkovskiy}\ and\ \citenamefont
  {Chakraborty}(2016)}]{PhysRevB.93.085145}%
  \BibitemOpen
  \bibfield  {author} {\bibinfo {author} {\bibfnamefont {P.~K.}\ \bibnamefont
  {Pyatkovskiy}}\ and\ \bibinfo {author} {\bibfnamefont {T.}~\bibnamefont
  {Chakraborty}},\ }\href@noop {} {\bibfield  {journal} {\bibinfo  {journal}
  {Phys. Rev. B}\ }\textbf {\bibinfo {volume} {93}},\ \bibinfo {pages} {085145}
  (\bibinfo {year} {2016})}\BibitemShut {NoStop}%
\bibitem [{\citenamefont {{Mawrie}}\ and\ \citenamefont
  {{Muralidharan}}()}]{ales}%
  \BibitemOpen
  \bibfield  {author} {\bibinfo {author} {\bibfnamefont {A.}~\bibnamefont
  {{Mawrie}}}\ and\ \bibinfo {author} {\bibfnamefont {B.}~\bibnamefont
  {{Muralidharan}}},\ }\href@noop {} {\bibinfo  {journal} {arXiv: 1810.05411
  [cond-mat.mes-hall]}\ }\BibitemShut {NoStop}%
\bibitem [{\citenamefont {Sriluckshmy}\ \emph {et~al.}(2018)\citenamefont
  {Sriluckshmy}, \citenamefont {Saha},\ and\ \citenamefont
  {Moessner}}]{sriluckshmy2018interplay}%
  \BibitemOpen
\bibfield  {journal} {  }\bibfield  {author} {\bibinfo {author} {\bibfnamefont
  {P.~V.}\ \bibnamefont {Sriluckshmy}}, \bibinfo {author} {\bibfnamefont
  {K.}~\bibnamefont {Saha}}, \ and\ \bibinfo {author} {\bibfnamefont
  {R.}~\bibnamefont {Moessner}},\ }\href@noop {} {\bibfield  {journal}
  {\bibinfo  {journal} {Phys. Rev. B}\ }\textbf {\bibinfo {volume} {97}},\
  \bibinfo {pages} {024204} (\bibinfo {year} {2018})}\BibitemShut {NoStop}%
\bibitem [{\citenamefont {Wang}(2018)}]{interaction}%
  \BibitemOpen
  \bibfield  {author} {\bibinfo {author} {\bibfnamefont {J.}~\bibnamefont
  {Wang}},\ }\href@noop {} {\bibfield  {journal} {\bibinfo  {journal} {J.
  Phys.: Condens. Matter}\ }\textbf {\bibinfo {volume} {30}},\ \bibinfo {pages}
  {125401} (\bibinfo {year} {2018})}\BibitemShut {NoStop}%
\bibitem [{\citenamefont {Saha}\ \emph {et~al.}(2017)\citenamefont {Saha},
  \citenamefont {Nandkishore},\ and\ \citenamefont
  {Parameswaran}}]{PhysRevB.96.045424}%
  \BibitemOpen
  \bibfield  {author} {\bibinfo {author} {\bibfnamefont {K.}~\bibnamefont
  {Saha}}, \bibinfo {author} {\bibfnamefont {R.}~\bibnamefont {Nandkishore}}, \
  and\ \bibinfo {author} {\bibfnamefont {S.~A.}\ \bibnamefont {Parameswaran}},\
  }\href@noop {} {\bibfield  {journal} {\bibinfo  {journal} {Phys. Rev. B}\
  }\textbf {\bibinfo {volume} {96}},\ \bibinfo {pages} {045424} (\bibinfo
  {year} {2017})}\BibitemShut {NoStop}%
\bibitem [{\citenamefont {{Pena-Benitez}}\ \emph {et~al.}()\citenamefont
  {{Pena-Benitez}}, \citenamefont {{Saha}},\ and\ \citenamefont
  {{Surowka}}}]{kush_arxiv}%
  \BibitemOpen
  \bibfield  {author} {\bibinfo {author} {\bibfnamefont {F.}~\bibnamefont
  {{Pena-Benitez}}}, \bibinfo {author} {\bibfnamefont {K.}~\bibnamefont
  {{Saha}}}, \ and\ \bibinfo {author} {\bibfnamefont {P.}~\bibnamefont
  {{Surowka}}},\ }\href@noop {} {\bibinfo  {journal} {arXiv: 1805.09827
  [cond-mat.str-el]}\ }\BibitemShut {NoStop}%
\bibitem [{\citenamefont {Li}\ and\ \citenamefont
  {Reichl}(1999)}]{PhysRevB.60.15732}%
  \BibitemOpen
\bibfield  {journal} {  }\bibfield  {author} {\bibinfo {author} {\bibfnamefont
  {W.}~\bibnamefont {Li}}\ and\ \bibinfo {author} {\bibfnamefont {L.~E.}\
  \bibnamefont {Reichl}},\ }\href@noop {} {\bibfield  {journal} {\bibinfo
  {journal} {Phys. Rev. B}\ }\textbf {\bibinfo {volume} {60}},\ \bibinfo
  {pages} {15732} (\bibinfo {year} {1999})}\BibitemShut {NoStop}%
\bibitem [{\citenamefont {Atteia}\ \emph {et~al.}(2017)\citenamefont {Atteia},
  \citenamefont {Bardarson},\ and\ \citenamefont
  {Cayssol}}]{PhysRevB.96.245404}%
  \BibitemOpen
  \bibfield  {author} {\bibinfo {author} {\bibfnamefont {J.}~\bibnamefont
  {Atteia}}, \bibinfo {author} {\bibfnamefont {J.~H.}\ \bibnamefont
  {Bardarson}}, \ and\ \bibinfo {author} {\bibfnamefont {J.}~\bibnamefont
  {Cayssol}},\ }\href@noop {} {\bibfield  {journal} {\bibinfo  {journal} {Phys.
  Rev. B}\ }\textbf {\bibinfo {volume} {96}},\ \bibinfo {pages} {245404}
  (\bibinfo {year} {2017})}\BibitemShut {NoStop}%
\bibitem [{\citenamefont {Eckardt}(2017)}]{RevModPhys.89.011004}%
  \BibitemOpen
  \bibfield  {author} {\bibinfo {author} {\bibfnamefont {A.}~\bibnamefont
  {Eckardt}},\ }\href@noop {} {\bibfield  {journal} {\bibinfo  {journal} {Rev.
  Mod. Phys.}\ }\textbf {\bibinfo {volume} {89}},\ \bibinfo {pages} {011004}
  (\bibinfo {year} {2017})}\BibitemShut {NoStop}%
\bibitem [{\citenamefont {Dutta}\ \emph {et~al.}(2013)\citenamefont {Dutta},
  \citenamefont {Maiti},\ and\ \citenamefont {Karmakar}}]{dutta}%
  \BibitemOpen
  \bibfield  {author} {\bibinfo {author} {\bibfnamefont {P.}~\bibnamefont
  {Dutta}}, \bibinfo {author} {\bibfnamefont {S.~K.}\ \bibnamefont {Maiti}}, \
  and\ \bibinfo {author} {\bibfnamefont {S.}~\bibnamefont {Karmakar}},\
  }\href@noop {} {\bibfield  {journal} {\bibinfo  {journal} {J. Appl. Phys.}\
  }\textbf {\bibinfo {volume} {114}},\ \bibinfo {pages} {034306} (\bibinfo
  {year} {2013})}\BibitemShut {NoStop}%
\bibitem [{\citenamefont {Shakouri}\ \emph {et~al.}(2015)\citenamefont
  {Shakouri}, \citenamefont {Simchi}, \citenamefont {Esmaeilzadeh},
  \citenamefont {Mazidabadi},\ and\ \citenamefont
  {Peeters}}]{PhysRevB.92.035413}%
  \BibitemOpen
  \bibfield  {author} {\bibinfo {author} {\bibfnamefont {K.}~\bibnamefont
  {Shakouri}}, \bibinfo {author} {\bibfnamefont {H.}~\bibnamefont {Simchi}},
  \bibinfo {author} {\bibfnamefont {M.}~\bibnamefont {Esmaeilzadeh}}, \bibinfo
  {author} {\bibfnamefont {H.}~\bibnamefont {Mazidabadi}}, \ and\ \bibinfo
  {author} {\bibfnamefont {F.~M.}\ \bibnamefont {Peeters}},\ }\href@noop {}
  {\bibfield  {journal} {\bibinfo  {journal} {Phys. Rev. B}\ }\textbf {\bibinfo
  {volume} {92}},\ \bibinfo {pages} {035413} (\bibinfo {year}
  {2015})}\BibitemShut {NoStop}%
\bibitem [{\citenamefont {Castro~Neto}\ \emph {et~al.}(2009)\citenamefont
  {Castro~Neto}, \citenamefont {Guinea}, \citenamefont {Peres}, \citenamefont
  {Novoselov},\ and\ \citenamefont {Geim}}]{RevModPhys.81.109}%
  \BibitemOpen
  \bibfield  {author} {\bibinfo {author} {\bibfnamefont {A.~H.}\ \bibnamefont
  {Castro~Neto}}, \bibinfo {author} {\bibfnamefont {F.}~\bibnamefont {Guinea}},
  \bibinfo {author} {\bibfnamefont {N.~M.~R.}\ \bibnamefont {Peres}}, \bibinfo
  {author} {\bibfnamefont {K.~S.}\ \bibnamefont {Novoselov}}, \ and\ \bibinfo
  {author} {\bibfnamefont {A.~K.}\ \bibnamefont {Geim}},\ }\href@noop {}
  {\bibfield  {journal} {\bibinfo  {journal} {Rev. Mod. Phys.}\ }\textbf
  {\bibinfo {volume} {81}},\ \bibinfo {pages} {109} (\bibinfo {year}
  {2009})}\BibitemShut {NoStop}%
\bibitem [{\citenamefont {Kohler}\ \emph {et~al.}(2005)\citenamefont {Kohler},
  \citenamefont {Lehmann},\ and\ \citenamefont
  {H{\"a}nggi}}]{kohler2005driven}%
  \BibitemOpen
  \bibfield  {author} {\bibinfo {author} {\bibfnamefont {S.}~\bibnamefont
  {Kohler}}, \bibinfo {author} {\bibfnamefont {J.}~\bibnamefont {Lehmann}}, \
  and\ \bibinfo {author} {\bibfnamefont {P.}~\bibnamefont {H{\"a}nggi}},\
  }\href@noop {} {\bibfield  {journal} {\bibinfo  {journal} {Phys. Rep.}\
  }\textbf {\bibinfo {volume} {406}},\ \bibinfo {pages} {379} (\bibinfo {year}
  {2005})}\BibitemShut {NoStop}%
\bibitem [{\citenamefont {Foa~Torres}(2005)}]{PhysRevB.72.245339}%
  \BibitemOpen
  \bibfield  {author} {\bibinfo {author} {\bibfnamefont {L.~E.~F.}\
  \bibnamefont {Foa~Torres}},\ }\href@noop {} {\bibfield  {journal} {\bibinfo
  {journal} {Phys. Rev. B}\ }\textbf {\bibinfo {volume} {72}},\ \bibinfo
  {pages} {245339} (\bibinfo {year} {2005})}\BibitemShut {NoStop}%
\end{thebibliography}%
\end{document}